\documentclass[preprint]{aastex61}
\usepackage{epstopdf}
\usepackage[toc]{appendix}
\usepackage{color}
\usepackage{lineno}
\received{\today}
\submitjournal{ApJ}

\shorttitle{Solar wind discontinuity}
\shortauthors{Kropotina et al.}

\begin{document}

\title{Solar wind discontinuity transformation at the bow shock}  

\correspondingauthor{Julia A. Kropotina}
\email{juliett.k@gmail.com}
\correspondingauthor{Anton V. Artemyev}
\email{aartemyev@igpp.ucla.edu}

\author{Julia A. Kropotina}
\affiliation{Ioffe Institute, St. Petersburg, 194021 Russia}

\author{Lee Webster}
\affiliation{Nyheim Plasma Institute, Drexel University, Camden, NJ, USA}

\author{Anton V. Artemyev}
\affiliation{Department of Earth, Planetary, and Space Sciences and Institute of Geophysics and Planetary Physics, \\University of California, Los Angeles, CA, USA}
\affiliation{Space Research Institute, RAS, Moscow, Russia}

\author{Andrei M. Bykov}
\affiliation{Ioffe Institute, St. Petersburg, Russia}

\author{Dmitri L. Vainchtein}
\affiliation{Nyheim Plasma Institute, Drexel University, Camden, NJ, USA}
\affiliation{Space Research Institute, RAS, Moscow, Russia}

\author{Ivan Y. Vasko}
\affiliation{Space Sciences Laboratory, University of California, Berkeley, CA, USA}
\affiliation{Space Research Institute, RAS, Moscow, Russia}

\begin{abstract}
Solar wind plasma at the Earth’s orbit carries transient magnetic field structures including discontinuities. Their interaction with the Earth’s bow shock can significantly alter discontinuity configuration and stability. We investigate such an interaction for the most widespread type of solar wind discontinuities -- rotational discontinuities (RDs). We use a set of {\it in situ} multispacecraft observations and perform kinetic hybrid simulations. We focus on the RD current density amplification that may lead to magnetic reconnection. We show that the amplification can be as high as two orders of magnitude and is mainly governed by three processes: the transverse magnetic field compression, global thinning of RD, and interaction of RD with low-frequency electromagnetic waves in the magnetosheath, downstream of the bow shock. The first factor is found to substantially exceed simple hydrodynamic predictions in most observed cases, the second effect has a rather moderate impact, while the third causes strong oscillations of the current density. We show that the presence of accelerated particles in the bow shock precursor highly boosts the current density amplification, making the postshock magnetic reconnection more probable. The pool of accelerated particles strongly affects the interaction of RDs with the Earth's bow shock, as it is demonstrated by observational data analysis and hybrid code simulations.  Thus, shocks should be distinguished not by the inclination angle, but rather by the presence of foreshocks populated with shock reflected particles. Plasma processes in the RD-shock interaction affect magnetic structures and turbulence in the Earth's magnetosphere and may have implications for the processes in  astrophysics.
\end{abstract}


\section{Introduction}
Solar wind flow carries a wide range of magnetic field structures (solar wind transients) that are large-amplitude Alfv\'{e}n waves, rotational (RD) and tangential discontinuities (TD, current sheets), and interplanetary shock waves \citep{Smith73:I, Smith73:II,Neugebauer84, Neugebauer06,Tsurutani&Ho99, Vasquez07, Greco09:pre, Greco09}. These structures significantly contribute to the magnetic fluctuation spectrum \citep{Borovsky10:solarwind}. A dissipation of magnetic field contained in solar wind transients is believed to support proton and electron heating \citep{Osman11:solarwind, Osman12:solarwind} and acceleration \citep{Tessein13}. This dissipation can be due to magnetic field line reconnection that is driven by internal or externally driven instabilities of strong currents carried by solar wind transients \citep{Phan06:reconnection, Gosling12,Zank14, Drake20}, e.g. solar wind RDs and TDs, that are plane layers of intense currents \citep{Servidio15, Podesta17, Podesta&Roytershteyn17, Greco16}. Solar wind discontinuities are generated as boundaries of plasma flows with different properties \citep{Servidio11, Servidio11:npg} or result from the nonlinear evolution of Alfv\'{e}n waves \citep{Medvedev97:pop, Medvedev97:prl, Vasquez&Hollweg99, Vasquez&Hollweg01}. RDs can be observed close to the Sun \citep{Phan20, Larosa20:arXiv, Krasnoselskikh20}, and they propagate to the Earth's orbit \citep[e.g.,][]{Vasquez07, Artemyev19:jgr:solarwind} and beyond \citep[e.g.,][]{Soding01, Artemyev18:apj}. Thus, RDs are supposed to be quite stable to internal instabilities \citep[as was confirmed by numerical models, see][]{Omidi92, Karimabadi95, Lin09:rotational_discontinuities}, and an external driver is often required to trigger magnetic reconnection, energy dissipation, and plasma heating. The most straightforward of such drivers are the interactions of RDs with each other or with larger-scale magnetic field structures, like planetary bow shocks \citep{Nakanotani20}.

The importance of geomagnetic perturbations induced by the interaction of solar wind discontinuities with the Earth's bow has been widely acknowledged, \citep{Tsurutani11:jastp, Zong&Zhang11, Farrugia13}. These perturbations were detected in the foreshock \citep{Lin97, Turner13:foreshock, Liu15:foreshock, An20:apj} and downstream of the bow shock, in the magnetosheath \citep{Archer12, Plaschke18}. Such perturbations are an effective source of energetic particles upstream of the bow shock \citep[e.g.,][]{Turner18:nature, Liu19:foreshock}. Propagation of these perturbations in the magnetosheath \citep[e.g.,][]{Farrugia95, Wang20:discontinuity} and their interaction with the boundary of the Earth's magnetosphere, the magnetopause, drive large-scale disturbances of the magnetosphere magnetic field and affect space weather \citep[e.g.,][]{Kokubun77, Tsurutani11:jastp, Korotova12,Turc15}.

Although observed {\it in situ} mostly in the Earth's magnetosphere, the interaction of discontinuities with shock waves is also a subject of theoretical and numerical investigations with application to astrophysical systems. One of the literally brightest examples is the pulsar magnetospheres \citep{Kennel&Coroniti84, Arons12}, where discontinuities are the natural part of the stripped pulsar wind \citep{Bogovalov99}, and their interaction with shock waves is believed to be responsible for charged particle acceleration and pulsar flares \citep[e.g.,][]{Sironi&Spitkovsky11, Arons12}. Moreover, a supersonically moving pulsar produces an extended bow shock pulsar wind nebula with a distinctively hard power-law spectra of the X-ray radiation observed by Chandra X-ray observatory \citep[see, e.g.][]{Posselt17,AB17,AB19}. Therefore, investigation of the solar wind discontinuity interaction with the Earth's bow shock can help in constructing realistic astrophysical models.

In the framework of a single-fluid MHD approach, the interaction of a discontinuity with the bow shock can be considered analytically as an interaction of two MHD discontinuities \cite[e.g.,][]{Volk&Auer74, Pushkar10}. However, the most interesting results that go beyond the analytical approximation can be obtained from numerical simulations and spacecraft observations. For the shock-shock interaction (when a solar wind discontinuity is an interplanetary shock wave interacting with the Earth's bow shock), large-scale MHD simulations and spacecraft observations suggest the generation of a set of {\it secondary} discontinuities \citep[e.g.,][]{Koval06, Samsonov07, Goncharov15} and deceleration of the interplanetary shock after the bow shock crossing \citep{Koval05b}. Energetic particles trapped between the interplanetary shock and the bow shock can be effectively accelerated \citep{Roth&Bale06, Hietala12:apj}. This acceleration was observed in the heliosphere, but it has a clear analogy in astrophysical systems \citep[e.g.,][]{Bykov13}. However, the occurrence rate of interplanetary shocks is much smaller than the occurrence rate of TDs and RDs \citep{Vasquez07, Soding01}.

Most of observations of solar wind discontinuities show that such solar wind transients share properties of both TDs and RDs \citep{Neugebauer84, Neugebauer06, Artemyev19:grl:solarwind, Newman20}, and a separation of these two types is possible only for case studies, when the local coordinate system of a discontinuity can be well-defined. TDs are plasma boundaries with significant diamagnetic currents \citep{deKeyser96, deKeyser97,Roth96, Neukirch20, Neukirch20:jpp}, which makes such discontinuities unstable to magnetic reconnection \citep{Swisdak03, Phan10} and they decay into pairs of intermediate shocks or RDs. Therefore, observational evidences of a TD interaction with the bow shock are quite limited \citep[e.g.,][]{Maynard07, Maynard08, Keika09}. Simulations suggest that such interaction should result in a generation of mesoscale dynamical structures in the magnetosheath and foreshock, e.g. plasma bubbles\citep{Lin97, An20:apj}, density cavities \citep{Omidi10}, and hot flow anomalies \citep{Lin02:HFA}. Much more stable RDs seem to dominate the statistics of solar wind transients \citep{Vasquez07, Neugebauer06, Artemyev19:jgr:solarwind}.

Resistive MHD simulations of RD interaction with the bow shock show that RDs transform (or decay) to slow and intermediate shocks downstream of the bow shock \citep{Yan&Lee95, Lin96:discontinuity, Cable&Lin98}. These shocks are boundaries of the plasma pressure pulse, the density and pressure enhancement that is also well-documented in observations of the RD interaction with the bow shock \citep[e.g.,][]{Hubert&Harvey00, Farrugia18}. Hybrid simulations, which treat ions as particles and electrons as a fluid, confirm these results but also show that intermediate shocks are replaced by RDs \citep{Lin96:discontinuity, Lin96:pressure_pulse}. Such transformation can be accompanied by the ion acceleration \citep{Farrugia18} and an increase of the tangential magnetic field of RDs \citep{Lin09:rotational_discontinuities}.

To investigate effects of RD interaction with the bow shock \citep{Lin96:discontinuity, Lin09:rotational_discontinuities} we should compare RD characteristics in the upstream and downstream. Indeed, RDs stably propagate in the solar wind \citep{Soding01, Artemyev18:apj, Artemyev19:grl:solarwind} and store a significant amount of magnetic field energy \citep{Borovsky10:solarwind}. The bow shock crossing may alter the RD properties (e.g., increase the current density or form a plasma pressure gradient across the discontinuity) and make RDs unstable to magnetic reconnection. This would explain a large number of observational reports about magnetic reconnection and charged particle acceleration in the magnetosheath \citep{Retino07:nat, Phan18:nature, Gingell20, Wang20:discontinuity, Bessho20}.
We emphasise the role of accelerated particles in the RD-shock interaction and the current density amplification. In a recent paper, \citep{Guo21}, the authors described 3D global hybrid simulations of the RD-shock interaction and showed that the upstream turbulence at a quasi-parallel bow shock made the RD unstable to magnetic reconnections. Those results are consistent with our findings.

Comparison of the kinetic properties of RDs upstream and downstream of the bow shock requires multispacecraft observations that would include monitoring the bow shock configuration and probing RDs at different locations relative to the bow shock. Moreover, such multispacecraft observations should be supplemented and supported by numerical simulations (preferably using hybrid codes that resolve ion kinetics). In this study, we combine observations of multispacecraft THEMIS/ARTEMIS missions \citep{Angelopoulos08:ssr, Angelopoulos11:ARTEMIS} and hybrid simulations \citep{Kropotina2018, Kropotina2019, Kropotina2018PAN, Kropotina2019conf, Kropotina2020} to investigate the evolution of RD properties due to the Earth's bow shock crossings. We distinguish three current amplification mechanisms during the bow shock transition. First, the RD can be thinned while crossing the shock precursor (if it exists). Second, the transverse magnetic field is amplified due to a simple shock compression (actually, the observed compression ratio far exceeds the Hugoniot prediction, while the simulated compression ratio is only slightly higher). The third effect is due to the RD interaction with the small-scale postshock magnetic turbulence. This leads to the complicated RD structure embedding strong alternating current oscillations.

The paper consists of an observational part and a description of the simulations. Section \ref{sec:obs} includes an overview of the spacecraft dataset, selection criteria, several examples of the RD-shock interaction, and a brief summary of observed patterns. In Section \ref{sec:sim} we describe our hybrid code as well as the simulation setup and results. In Section \ref{sec:concl} we present a discussion and our conclusions.

\section{Spacecraft observations}
\label{sec:obs}
We analyzed measurements of THEMIS fluxgate magnetometers \citep[FGM;][]{Auster08:THEMIS} and electrostatic analyzers \citep[ESA;][]{McFadden08:THEMIS}. FGM provides magnetic field vector ${\bf B}$ (in the GSE coordinates), and we used spin-averaged (4s) data. ESA provides omnidirectional ion spectra for a $<30$ keV energy range, ion bulk flow speed ${\bf V}$, and electron density $n$ (see a discussion of the ESA accuracy of measurements in the solar wind in \citep{Artemyev18:jgr:report}) with a spin 4s resolution. Three THEMIS spacecraft move along elliptical orbits with the apogee about $\sim 12$ Earth radii, $R_E$, and regularly form a configuration with one of the spacecraft in the foreshock (upstream of the Earth's bow shock), and two remaining spacecrafts in the bow shock and magnetosheath (downstream). Such THEMIS configurations are often used to analyze foreshock transients \citep[e.g.,][]{Turner13:foreshock, Liu15:foreshock} and shock structure \citep[e.g.,][]{Pope19, Gedalin20:angeo}. Two ARTEMIS spacecraft are orbiting the Moon ($\sim 60R_E$ from the Earth and $\sim 50R_E$ from the subsolar point of the bow shock) and regularly observe a pristine solar wind upstream of the Earth's bow shock \citep[these spacecraft measurements are often used to analyze the solar wind transients, e.g. discontinuities and interplanetary shocks; see][]{Artemyev19:jgr:solarwind, Davis20, Zhou20:shock}.

\subsection{Selection criteria and methods}
We selected twenty events that all had similar spacecraft configurations and set of measurements:
\begin{itemize}
\item ARTEMIS spacecraft observed a solar wind discontinuity that is (a) sufficiently strong (the magnitude of magnetic field variations exceeds $1$ nT), (b) well distinguished (there are no discontinuities with similar magnetic field magnitudes $\pm 10$ minutes around), (c) rotational with a compressional magnetic field variation $<0.25$ nT;
\item one of THEMIS spacecraft observed the same discontinuity (the discontinuity identity is determined from the similarity of the overall magnetic field configuration and an approximate ARTEMIS/THEMIS propagation time) downstream of the bow shock;
\item another THEMIS spacecraft probed the bow shock structure and then observed the same discontinuity downstream of the bow shock.
\end{itemize}

The current density is one of the main properties of discontinuities; it determines the stability \citep{bookGaleev85:vol2, Kuznetsova95, KDQ05theory} and configuration of a discontinuity. To determine the current density, we transform the discontinuity magnetic field into a local coordinate system determined using the Minimum Variance Analysis (MVA); \citep{Sonnerup68}: vector ${\bf l}$ is along the most varying magnetic field component (the tangential magnetic field of the discontinuity), vector ${\bf n}$ is along the least varying magnetic field component (presumably along the discontinuity normal direction), and vector ${\bf m}={\bf n}\times{\bf l}$. The main current density component is $j_m=(1/\mu_0)\partial B_l/\partial r_n$, where $r_n=\int{V_n(t) \: dt}$ and $V_n={\bf V}\cdot {\bf n}$. However, single-spacecraft measurements are generally insufficient to distinguish between ${\bf n}$ and ${\bf m}$ vectors reasonably well \citep{Sergeev06, Knetter04, Horbury01}, and the plasma velocity along the actual normal can be significantly underestimated, resulting in unrealistically large $j_m$. Thus, instead of $r_n$ we use $r_{nm}=\int{\sqrt{V_n^2(t)+V_m^2(t)} \: dt}$, which may overestimate current sheet spatial scales, resulting in an underestimate of $j_m$, but not sensitive to uncertainties of vectors ${\bf m}$ and ${\bf n}$ \citep[see, e.g., discussion in ][]{Newman20}. The normal vectors for the selected bow shock crossings were determined using the methods based on the Rankine-Hugoniot conditions \citep{Vinas&Scudder86}. We also normalize $r_{nm}$ on the proton inertial length, $d_p=\sqrt{m_p/\mu_0\langle{n}\rangle e^2}$, with $\langle{n}\rangle$  being the density averaged over a discontinuity crossing.

Below, we give four examples of discontinuity-shock interactions, the first two showing quasi-parallel and quasi-perpendicular bow shock crossings by weak (small magnetic field magnitude) discontinuities, and the latter two by strong discontinuities.

\subsection{Example event \#1}
\begin{figure}
\centering
\includegraphics[width=1\textwidth]{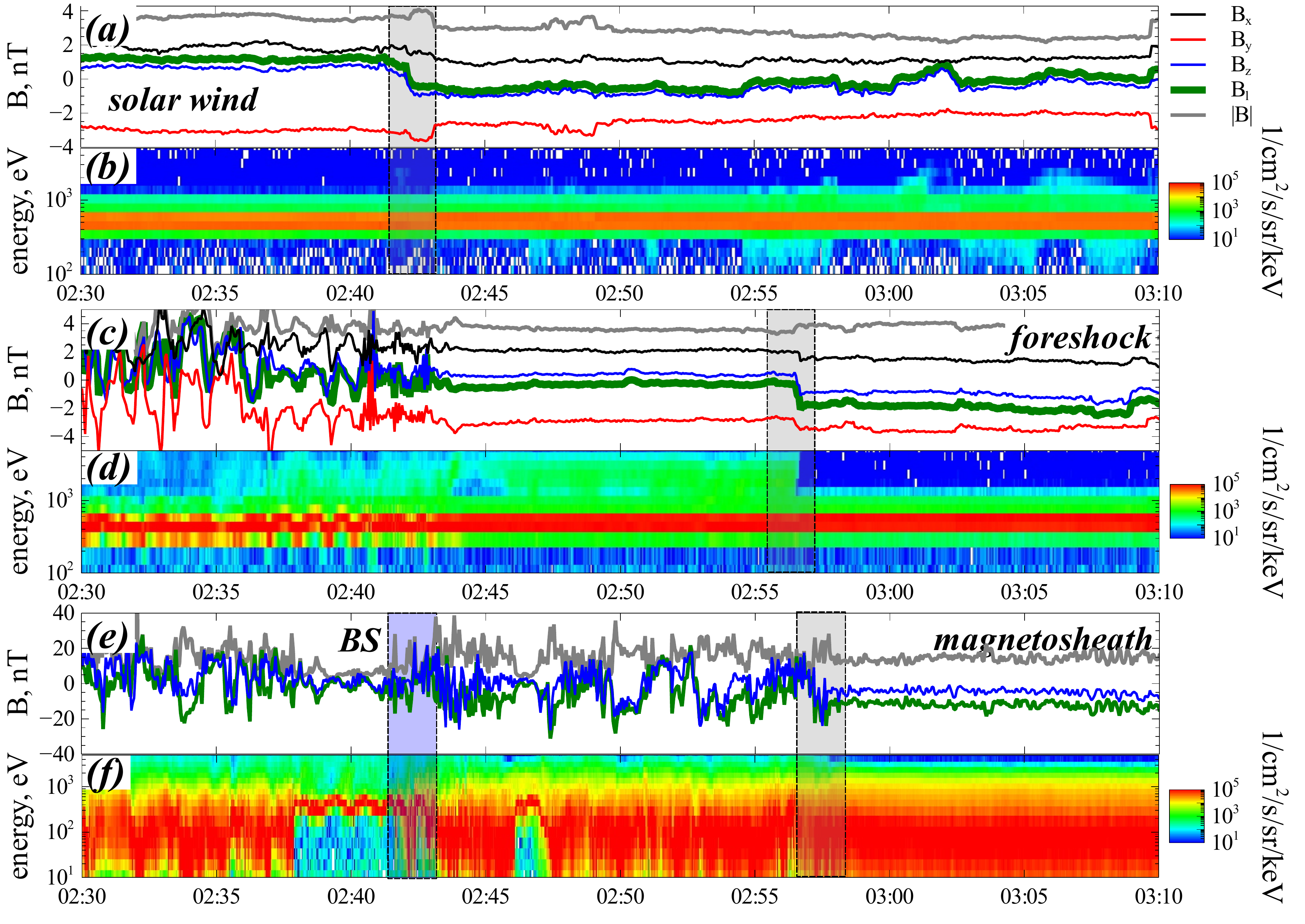}
\caption{\label{fig1} Observations of ARTEMIS P2 (originally THEMIS C; panels (a) and (b)), THEMIS D (panels (c) and(d)), and A (panels (e) and (f)) on 2019 December 3. ARTEMIS measurements are in the solar wind and THEMIS measurements are in the foreshock and in the magnetosheath. Discontinuity observation by each spacecraft is marked by the gray color. The bow shock (BS) crossing is marked by blue color in panels (e) and (f). Panels (a),(c), and (e) show magnetic field (three GSE components, magnitude, and the most varying component $B_l$). Local coordinate system used to plot $B_l$ has been constructed for magnetic field $\pm 1$ minutes around the discontinuity crossing. Panels (b), (d), and (f) show ion spectra for ARTEMIS and THEMIS observations.}
\end{figure}

Figure \ref{fig1} shows the first example event from our dataset: a weak discontinuity crosses a quasi-parallel ($\theta=23^\circ$) low Mach number ($M_A=2.8$) bow shock. Panels (a) and (b) show a discontinuity observed by ARTEMIS P1 in the pristine solar wind: a $B_z$ reversal is accompanied by a $B_y$ peak. The most varying magnetic field component, $B_l$, almost coincides with $B_z$, i.e. the normal direction to the discontinuity surface is in the $(x,y)$ plane. The solar wind ion spectrum does not show any variations right at the $B_z$ reversal. However, there is evidence of the presence of foreshock ions after this reversal: the spectrum shows ion populations with energy exceeding the solar wind energy. Thus, due to the magnetic field rotation at the discontinuity, the vector of the magnetic field may point from ARTEMIS to the bow shock, allowing the particles accelerated in vicinity of the bow shock to stream along the RD to the ARTEMIS location.

Panels (c) and (d) show the same discontinuity observed by THEMIS D in the ion foreshock region, just upstream of the bow shock. This discontinuity has the same magnetic field configuration ($B_l\approx B_z$) and it arrives to the THEMIS spacecraft from ARTEMIS within the expected time of solar wind travel, $\sim 10$ minutes. The discontinuity separates empty field lines and field lines connected to the bow shock and filled with high-energy reflected ions. Such reflected ions can be trapped by the discontinuity and modify its configuration.

Panels (e) and (f) shows observations of THEMIS A, which crossed the bow shock (at 02:42) and then probed the same discontinuity in the magnetosheath, downstream of the bow shock (at 02:57). The ion spectrum in the magnetosheath shows ion heating at the bow shock. The discontinuity in the magnetosheath keeps the same configuration (with $B_l\approx B_z$) as in the foreshock and the solar wind. In the magnetosheath, this discontinuity separates a region with high-amplitude magnetic field fluctuations from a region with a more laminar magnetic field. This effect will also be shown in hybrid simulations.

\begin{figure}
\centering
\includegraphics[width=1\textwidth]{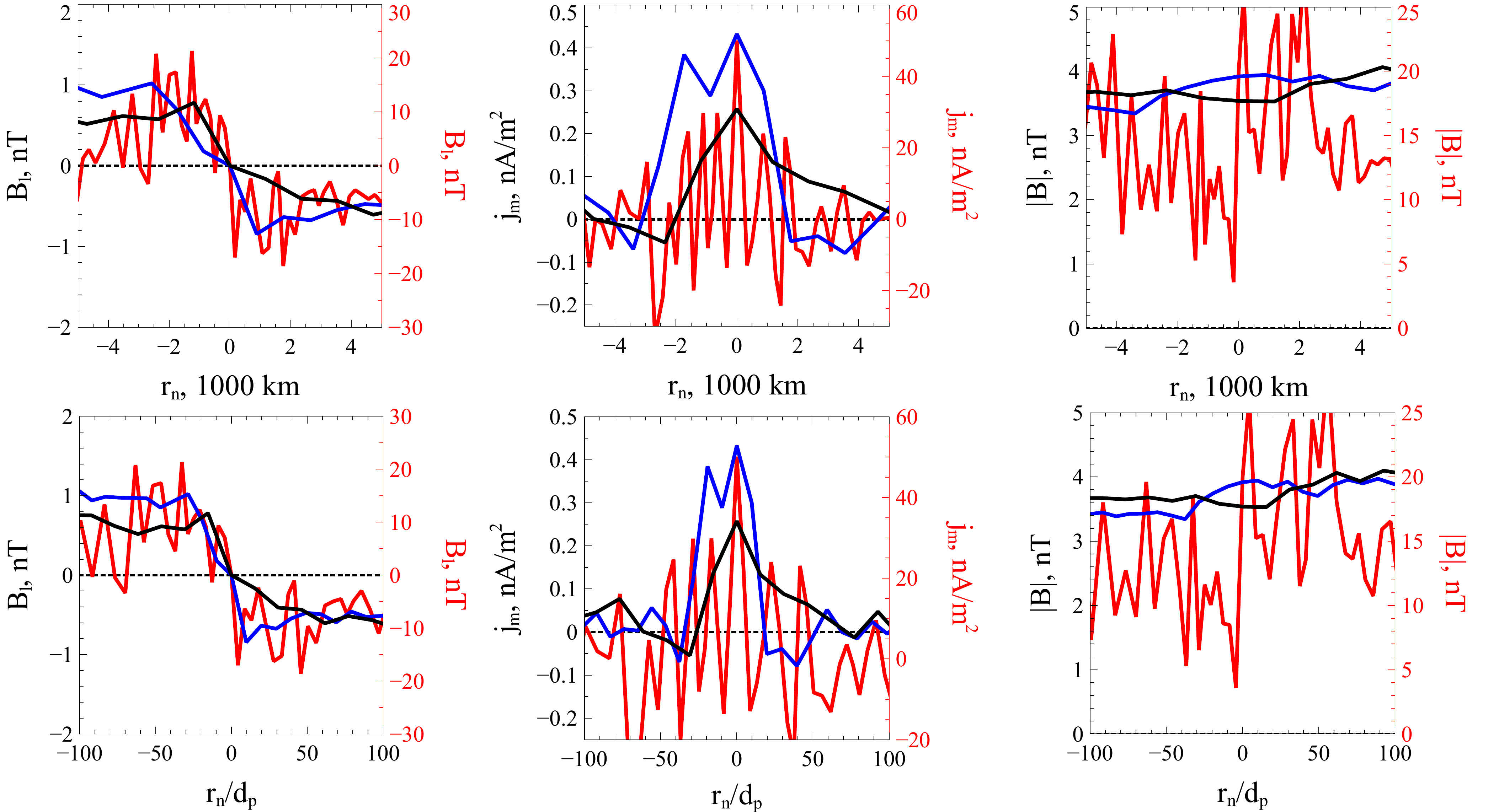}
\caption{\label{fig2} Magnetic field $B_l$, current density $j_m$, and magnetic field magnitude $|{\bf B}|$ for discontinuities from Figure~\ref{fig1}. The lines indicate observations in the solar wind (black), foreshock (blue), and magnetosheath (red). The spatial scale in the bottom row is normalized by the local proton inertial length, $d_p$.}
\end{figure}

Figure \ref{fig2} shows the magnetic field component $B_l$, current density $j_m=(1/\mu_0)\partial B_l/\partial r_n$, and magnetic field magnitude $|{\bf B}|$ for discontinuities from Figure \ref{fig1}. There is a clear increase of the current density magnitude and $B_l$ magnitude from the solar wind (where $\max B_l<1$ nT and $j_m\sim 0.2$ nA m$^{-2}$) to the foreshock (where $\max B_l \sim 1$ nT and $j_m\sim 0.4$ nA m$^{-2}$), and then to the magnetosheath (where $\max B_l\sim 15$ nT and $j_m\sim 30-40$ nA m$^{-2}$). Note that magnetic field fluctuations and waves in the magnetosheath significantly perturb $B_l$ and cause oscillations of $j_m$. These effects lead to strong intermittent local currents, which in turn may make the discontinuity unstable to magnetic reconnection \citep[e.g.,][]{Gingell20, Wang20:discontinuity, Bessho20}.

In the case of a smooth (single-scale) magnetic field profile, the RD thickness can be estimated as
\begin{equation}
\label{eq:L}
L\sim \frac{\max B_l - \min B_l}{2\mu_0 \max j_{m}}
\end{equation}
We found that $L \sim 2000$ km in the solar wind and $L \sim 1700$ km in the foreshock. In the magnetosheath, Eq.~(\ref{eq:L}) gives $L \sim 300$ km, but we should note that $j_{m}$ there can be contributed to wave-like magnetic field perturbations, and $L$  estimates some scale between discontinuity thickness and a much smaller wavelength of these perturbations. At the same time, the foreshock $B_l$ profile visually resembles the envelope of the oscillating magnetosheath $B_l$ profile, i.e. the discontinuity is thinned only while approaching the shock front. Remarkably, in terms of inertial lengths, the discontinuity thickness changes only slightly from the pristine solar wind to the magnetosheath (see the bottom panel in Figure~\ref{fig2}).

Note that we used a 4s resolution magnetic field because measurements with the higher available resolution, (1/5)s, are full of small-scale, large-amplitude (a fraction of $\max B_l$) fluctuations that contribute significantly to $\partial B_l/\partial r_n$ and mask the real current density profile. Thus, the absolute values of $j_m$ are effectively reduced (and correspondingly the values of $L$ are increased) tenfold by choosing a low magnetic field time resolution \citep[see discussion in][]{Artemyev18:apj}.

From Figure \ref{fig2}, we can estimate how different factors contribute to the observed two orders of magnitude current amplification. As we mentioned, estimates of discontinuity thinning give only a moderate amplification factor of $\sim 1.2$. At the same time, the transverse magnetic field $B_l$ is amplified by approximately a factor of $\sim 15$, which is quite large comparing to the usual Hugoniot prediction of $4$. Finally, the remaining factor of $\sim 10$ must be attributed to the observed downstream $B_l$ oscillations.

Right panels of Figure \ref{fig2} show the magnitude of magnetic field $|{\bf B}|$ around the discontinuity. Variations of $|{\bf B}|$ indicate compressible plasma fluctuations. We see that, in the solar wind and foreshock region, the discontinuity  is almost incompressible  ($|{\bf B}|\approx {\rm const}$), but the same discontinuity in the magnetosheath is characterized by a drop of $|{\bf B}|$ around the $j_m$ peak (at the discontinuity center). This drop is associated with a plasma pressure peak known as the pressure/density pulse \citep[see][]{Yan&Lee95, Lin96:discontinuity, Cable&Lin98}.

\subsection{Example event \#2}
\begin{figure}
\centering
\includegraphics[width=1\textwidth]{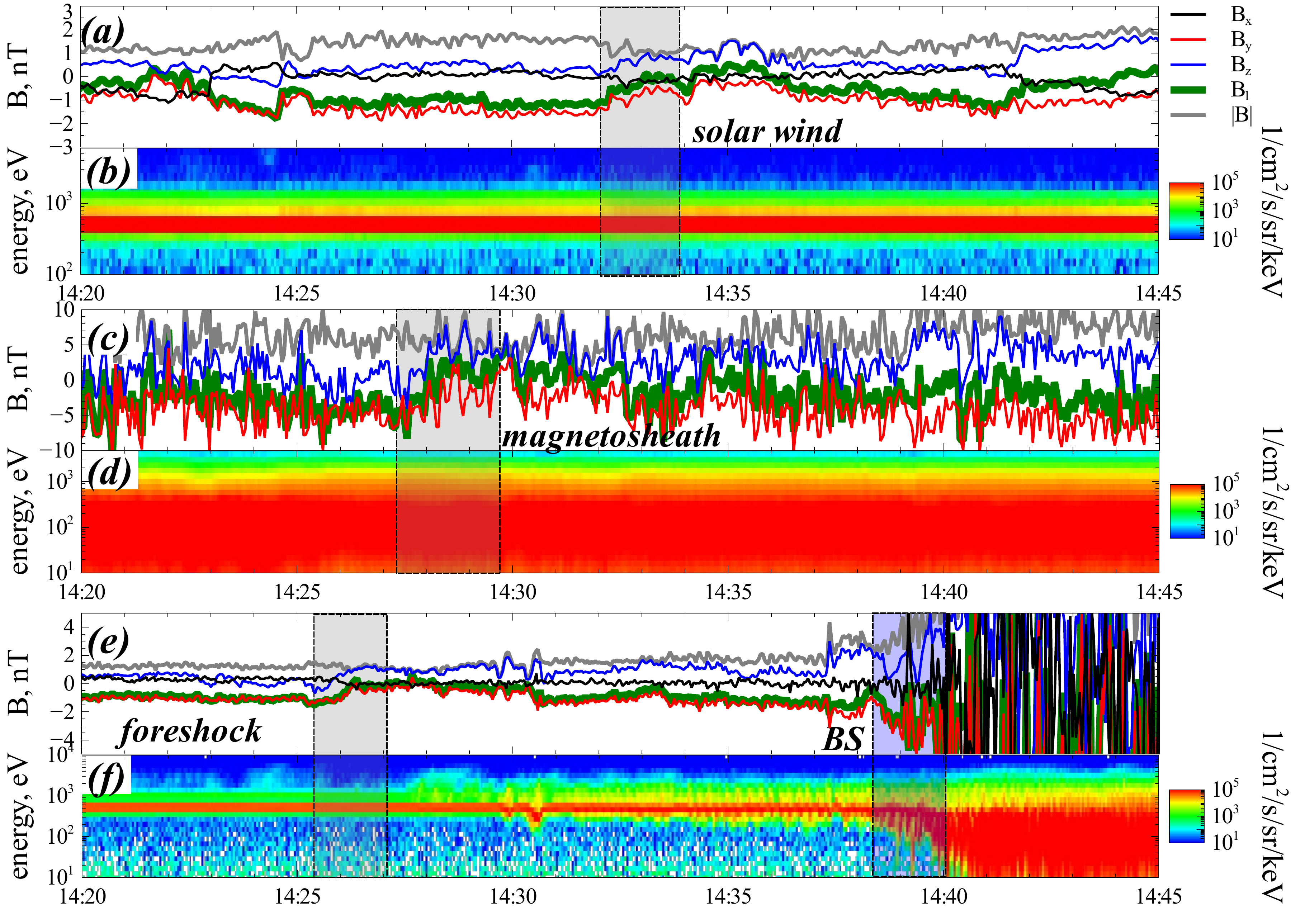}
\caption{\label{fig3} Observations of ARTEMIS P1 (originally THEMIS B), THEMIS D, and THEMIS A on 2019 December 5. The format is the same as in Figure~\ref{fig1}.}
\end{figure}

Figure \ref{fig3} shows the second example event from our dataset: a weak (small magnetic field magnitude) discontinuity crosses a quasi-perpendicular ($\theta=70^\circ$) high Mach number ($M_A=20$) bow shock. Panels (a) and (b) show the discontinuity observed by ARTEMIS P1 in the pristine solar wind: a $B_y$ reversal is accompanied by a $B_z$ peak. The most varying magnetic field component, $B_l$, almost coincides with $B_y$, i.e., the normal direction to the discontinuity surface is in the $(x,z)$ plane. Ion spectrum in the solar wind plasma does not show any accelerated ions arriving from the bow shock, which means that the discontinuity is not connected to the bow shock.

Panels (e) and (f) show the same discontinuity observed by THEMIS A in the ion foreshock region, just upstream of the bow shock that was crossed by THEMIS A fifteen minutes later (at 14:40). This discontinuity has the same magnetic field configuration ($B_l\approx B_y$). ARTEMIS spacecraft observed the discontinuity behind the terminator, i.e., the discontinuity first arrived to THEMIS A around the subsolar bow shock position and $\sim 10$ minutes later arrived to ARTEMIS. THEMIS A observations in the foreshock show the population of accelerated ions in the front vicinity, but the discontinuity has not entered this region yet.

Panels (c) and (d) show observations of THEMIS D probing discontinuity in the magnetosheath (as indicated by a high level of magnetic field fluctuations and wide energy ion spectrum), downstream of the bow shock (at 14:30). The discontinuity in the magnetosheath keeps the same configuration (with $B_l\approx B_y$) as in the foreshock and solar wind.

\begin{figure}
\centering
\includegraphics[width=1\textwidth]{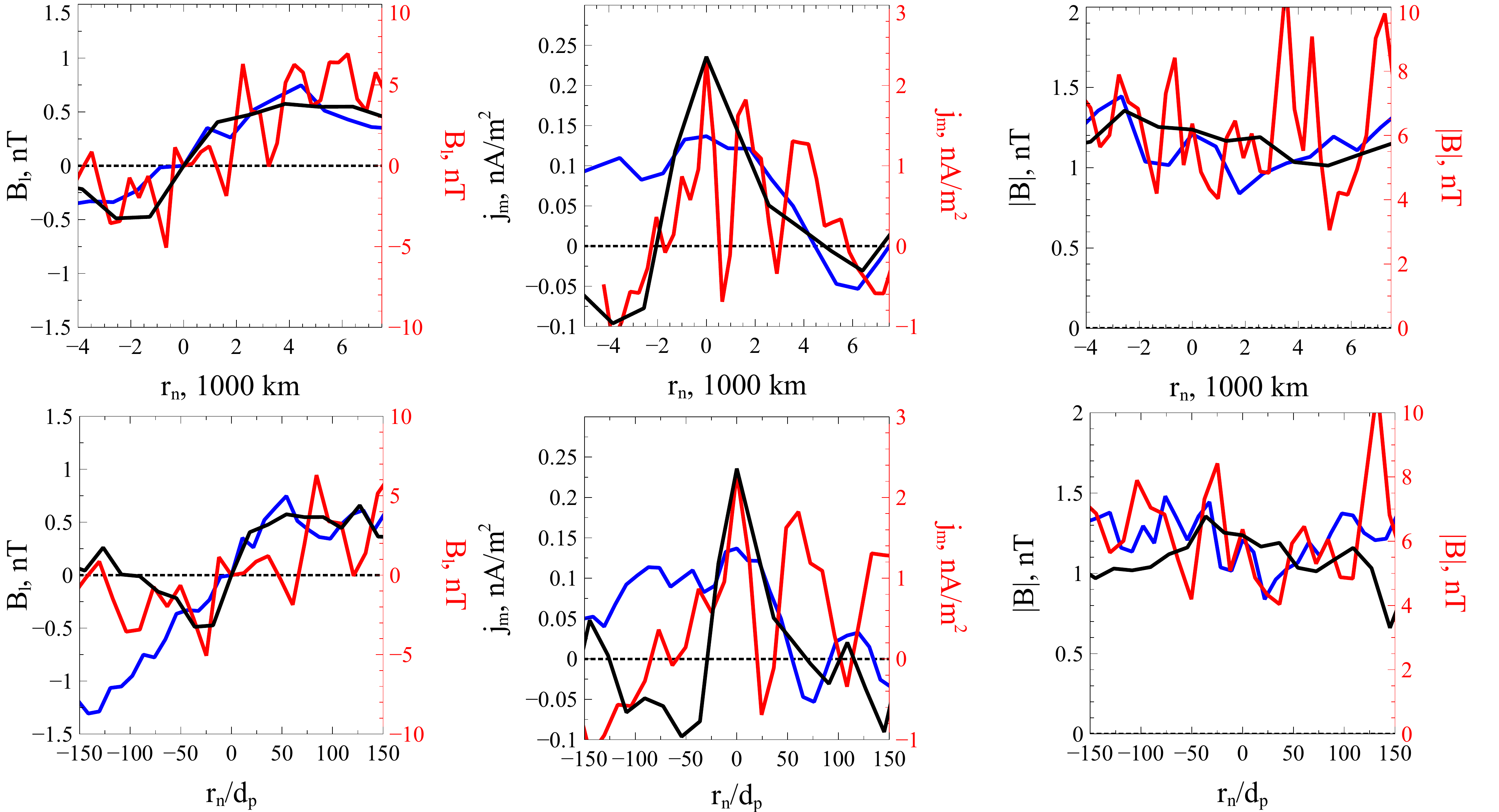}
\caption{\label{fig4} Magnetic field $B_l$, current density $j_m$, and magnetic field magnitude $|{\bf B}|$ for discontinuities from Figure~\ref{fig3}. The format is the same as in Figure~\ref{fig2}.}
\end{figure}

Figure \ref{fig4} shows $B_l$, $j_m$, and $|{\bf B}|$ for discontinuities from Figure \ref{fig3}. There are no significant differences between discontinuities observed in the solar wind (black) and foreshock (blue): lines for $B_l$ and $j_m$ are quite close for these two observations. This is likely because the discontinuity has not entered the region occupied by reflected particles yet (see the discussion of simulations results below) and was just moving through the unperturbed upstream solar wind. The discontinuity observed in the magnetosheath has the same $B_l(r_n)$ profile (though with more fluctuations), but $B_l$ increased from $0.5$ to $5$ nT. The corresponding increase of the current density is from $0.2$ to $2$ nA/m$^2$, i.e., the $\max B_l/j_m$ ratio does not change at the bow shock crossing. The discontinuity thickness $L \sim 1500\; {\rm km} \sim 15d_p$ remains the same, therefore the thin solar wind discontinuity does not become thinner due to the bow shock crossing (note that the thickness is overestimated due to usage of 4s resolution magnetic field). Discontinuities observed in the solar wind and magnetosheath are essentially incompressible with $|{\bf B}|\approx{\rm const}$, see the right panels in Figure~\ref{fig4}.


\subsection{Example event \#3}
\begin{figure}
\centering
\includegraphics[width=1\textwidth]{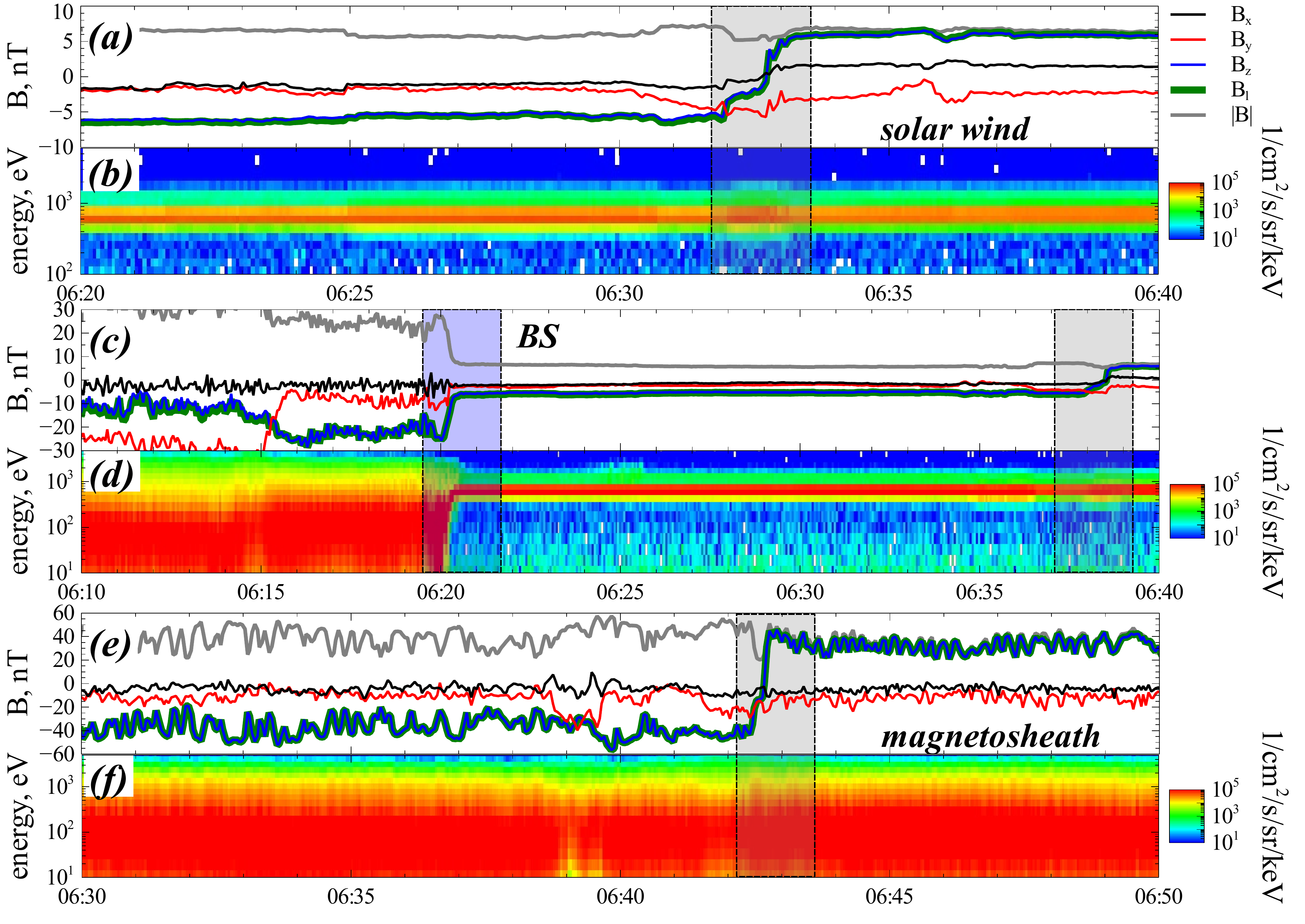}
\caption{\label{fig5} Observations of ARTEMIS P1 (originally THEMIS B), THEMIS D, and THEMIS A on 2018 October 13. The format is the same as in Figure \ref{fig1}.}
\end{figure}

Figure \ref{fig5} shows the third example event from our dataset: a strong (large magnetic field magnitude) discontinuity crosses a quasi-perpendicular ($\theta=105^\circ$) low Mach number ($M_A=3$) bow shock. Panels (a) and (b) show the discontinuity observed by ARTEMIS P1 in the pristine solar wind: a $B_z$ reversal is accompanied by a $B_y$ peak. The most varying magnetic field component, $B_l$, almost coincides with $B_z$, i.e. the normal direction to the discontinuity surface is in the $(x,y)$ plane.


\begin{figure}
\centering
\includegraphics[width=1\textwidth]{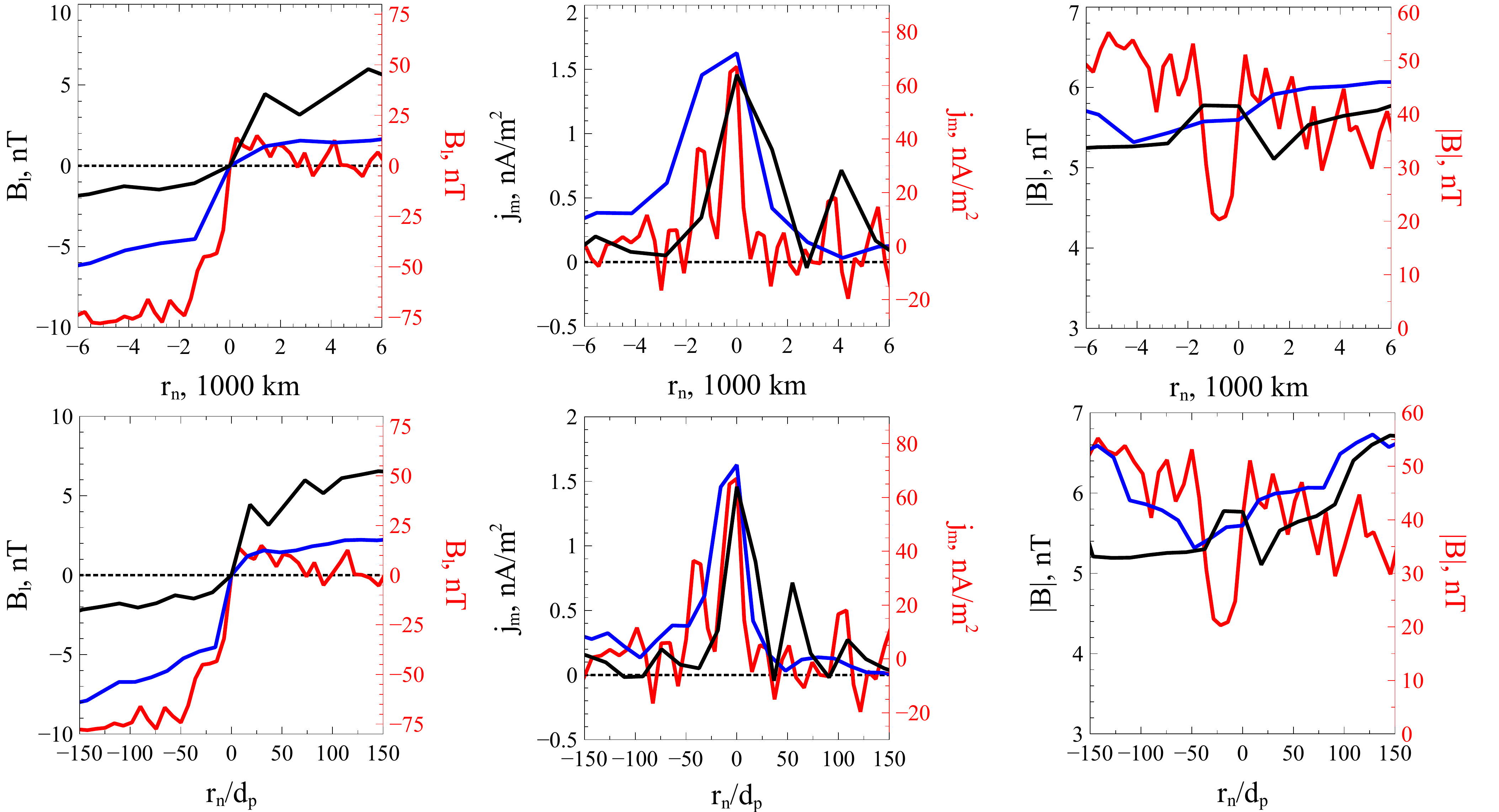}
\caption{\label{fig6} Magnetic field $B_l$, current density $j_m$, and magnetic field magnitude $|{\bf B}|$ for discontinuities from Figure~\ref{fig5}. The format is the same as in Figure~\ref{fig2}.}
\end{figure}

Panels (c) and (d) show the same discontinuity observed by THEMIS D in the foreshock region, just upstream of the bow shock that was crossed by this THEMIS D fifteen minutes before (at 06:20). This discontinuity has the same magnetic field configuration ($B_l\approx B_z$). ARTEMIS P2 and THEMIS D were almost at the same $x$ (Earth-Sun direction) coordinate and the $\sim 5$ minute delay of discontinuity observation between these two spacecraft is likely due to discontinuity orientation. THEMIS D observations in the foreshock show no reflected ions.

Panels (e) and (f) show observations of THEMIS D in the magnetosheath (as indicated by a high level of magnetic field fluctuations and an energy-wide ion spectrum), downstream of the bow shock (at 06:42). The discontinuity in the magnetosheath keeps the same configuration (with $B_l\approx B_y$) as in the foreshock and solar wind, but has much more evident magnetic field compression ($|{\bf B}|$ variation). Moreover, discontinuity in the magnetosheath separates regions with compressional magnetic fluctuations of different frequencies: after discontinuity crossing, THEMIS observes quasi-periodic $\sim 0.1$Hz magnetic fluctuations resembling mirror mode structures \citep{Soucek08}.

Figure \ref{fig6} shows $B_l$, $j_m$, and $|{\bf B}|$ for discontinuities from Figure \ref{fig5}. Although $B_l$ for discontinuities observed in the solar wind (black) and foreshock (blue) are not identical, these discontinuities have a similar current density magnitude, $j_m \sim 1.5$ nA m$^{-2}$, and thickness, $L \sim 1000\; {\rm km} \sim 10d_p$. After the bow shock crossing, $\max B_l$ increases from $7$ to $75$ nT, but the current density increase is even higher: from $1.5$ to $60$ nA m$^{-2}$. Such a current density increase is associated with the discontinuity thinning: from $L \sim 1000$ to $L \sim 250$ km. However, the discontinuity thinning does not reduce dimensionless thickness $L/d_p$. The right panels in Figure \ref{fig6} show that, after the bow shock crossing (upon entering the magnetosheath) a compressionless solar wind discontinuity with $|{\bf B}|\approx {\rm const}$ forms a strong $|{\bf B}|$ drop (associated with the plasma pressure peak); see also Figure \ref{fig2} and \citep{Yan&Lee95, Lin96:discontinuity, Cable&Lin98}.

\subsection{Example event \# 4}
\begin{figure}
\centering
\includegraphics[width=1\textwidth]{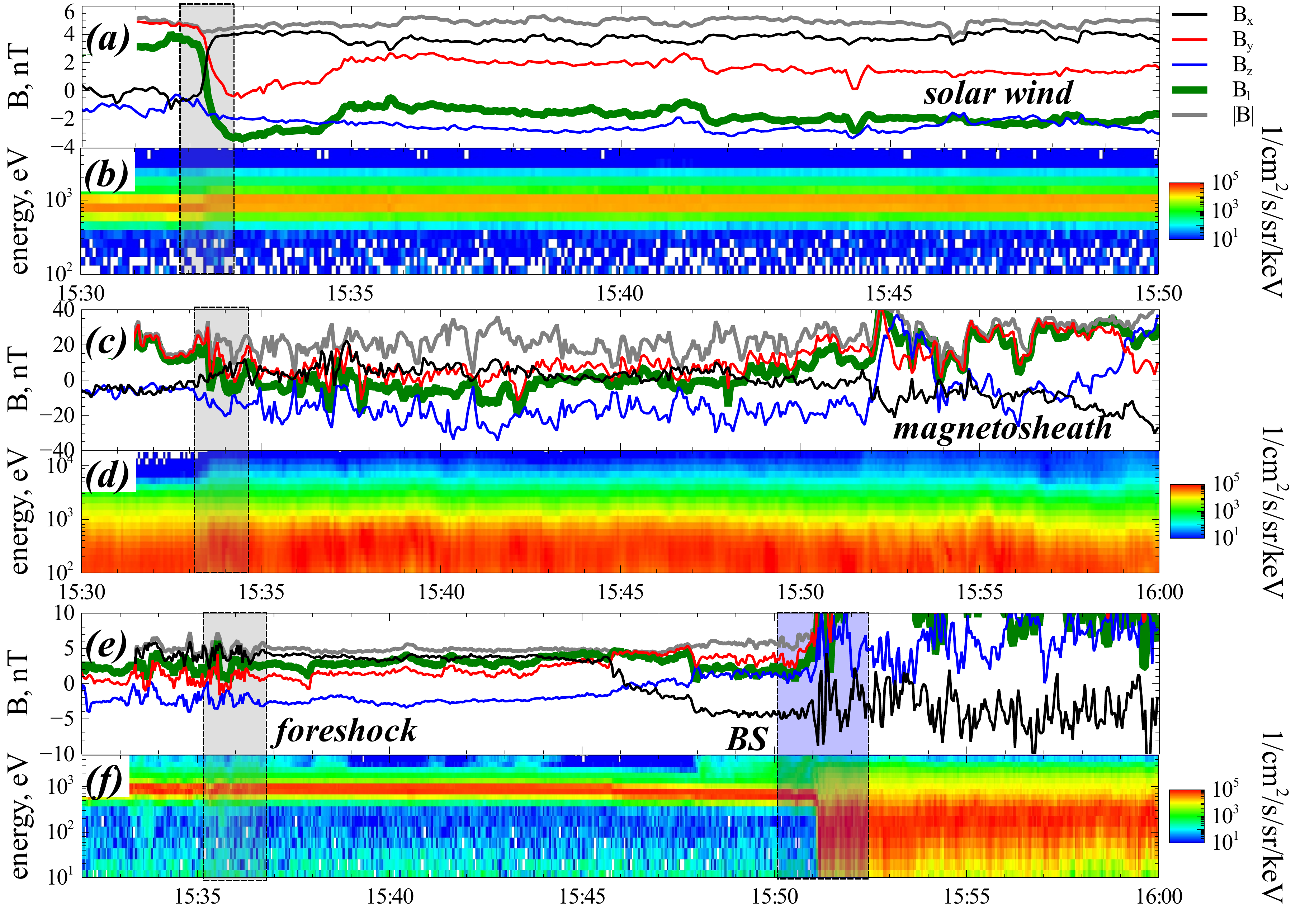}
\caption{\label{fig7} Observations of ARTEMIS P2 (originally THEMIS C), THEMIS D, and THEMIS A on 2019 November 5. The format is the same as in Figure~\ref{fig1}.}
\end{figure}

Figure \ref{fig7} shows the fourth example event from our dataset: a strong (large magnetic field magnitude) discontinuity crosses a quasi-parallel ($\theta=40^\circ$) low Mach number ($M_A=3$) bow shock. Panels (a) and (b) show the discontinuity observed by ARTEMIS P1 in the pristine solar wind: reversals of $B_x$ and $B_y$,  but almost constant $B_z$. The most varying magnetic field component, $B_l$, consists of $B_y$ and $B_x$, and the discontinuity normal direction is close to the $z$ direction. The ion spectrum in the solar wind plasma shows change of fluxes of the main (thermal) population, but no high-energy ions are observed, i.e, the discontinuity is likely disconnected from the bow shock.

\begin{figure}
\centering
\includegraphics[width=1\textwidth]{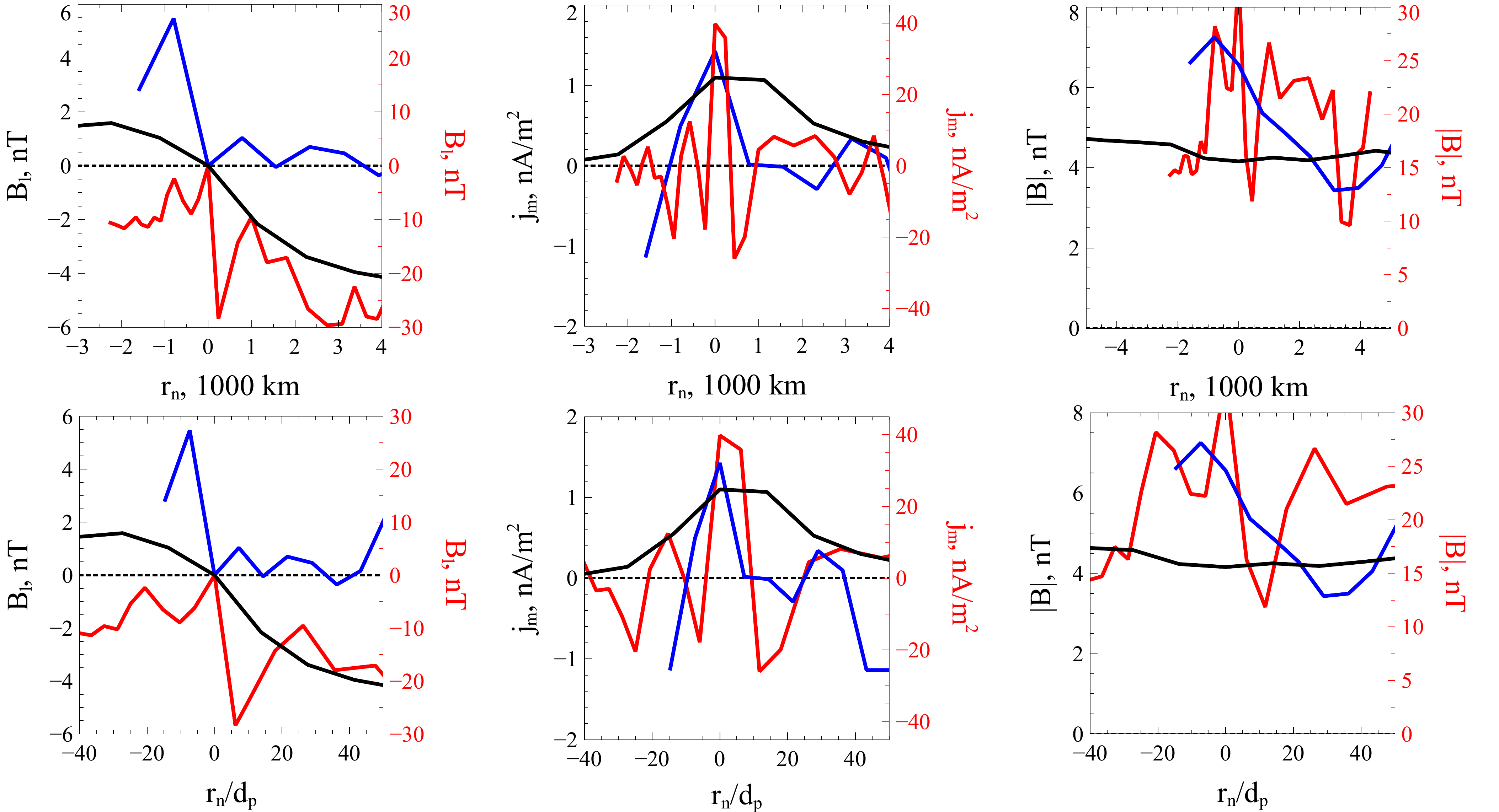}
\caption{\label{fig8} Magnetic field $B_l$, current density $j_m$, and magnetic field magnitude $|{\bf B}|$ for discontinuities from Figure~\ref{fig7}. The format is the same as in Figure~\ref{fig2}.}
\end{figure}

Panels (e) and (f) show the same discontinuity observed by THEMIS A in the foreshock region, just upstream of the bow shock that was crossed by THEMIS A fifteen minutes later (at 15:52). This discontinuity has the same magnetic field configuration as the solar wind discontinuity (same shapes of $B_x$ and $B_y$). The ARTEMIS and THEMIS spacecraft are well-separated along the $x$-coordinate (the Earth-Sun direction), and THEMIS A observed the discontinuity with a $\sim 5$ minute delay. THEMIS A observations in the foreshock show a large high-energy population of reflected ions around the discontinuity.

Panels (c) and (d) show observations of THEMIS D probing the same discontinuity (same $B_x$ and $B_y$ profiles) in the magnetosheath (visible from the high level of magnetic field fluctuations and energy wide ion spectrum), downstream of the bow shock. This discontinuity is associated with increase of energy of the main ion population.

Figure \ref{fig8} shows $B_l$, $j_m$, and $|{\bf B}|$ for discontinuities from Figure \ref{fig7}. There is a  clear discontinuity thinning from the solar wind $L\sim 2000$ km to the foreshock $L\sim 1000$ km, and further to the magnetosheath $L\sim 200$ km. This thinning is also seen in $j_m(r_n/d_p)$ profiles: $L\sim 20d_p$ in the solar wind, $L\sim 10d_p$ in the foreshock, and $L\sim 3d_p$ in the magnetosheath. The discontinuity thinning is accompanied by a $j_m$ increase from $1$ nA/m$^2$ in the solar wind to $40$ nA/m$^2$ in the magnetosheath; this increase is stronger than the $\max B_l$ increase from $4$ to $30$ nT. The left panels of Figure \ref{fig8} show that compressionless solar wind discontinuity with $|{\bf B}|\approx {\rm const}$ becomes quite compressional in the foreshock and in the magnetosheath. Comparison of Figures \ref{fig2},     \ref{fig4}, \ref{fig6}, and \ref{fig8} shows similar trends in current density increase, discontinuity thinning, and formation of a $|{\bf B}|$ minimum after the bow shock crossing. We summarize these trends in the next subsection.

\subsection{General trends}

\begin{figure}
\centering
\includegraphics[width=1\textwidth]{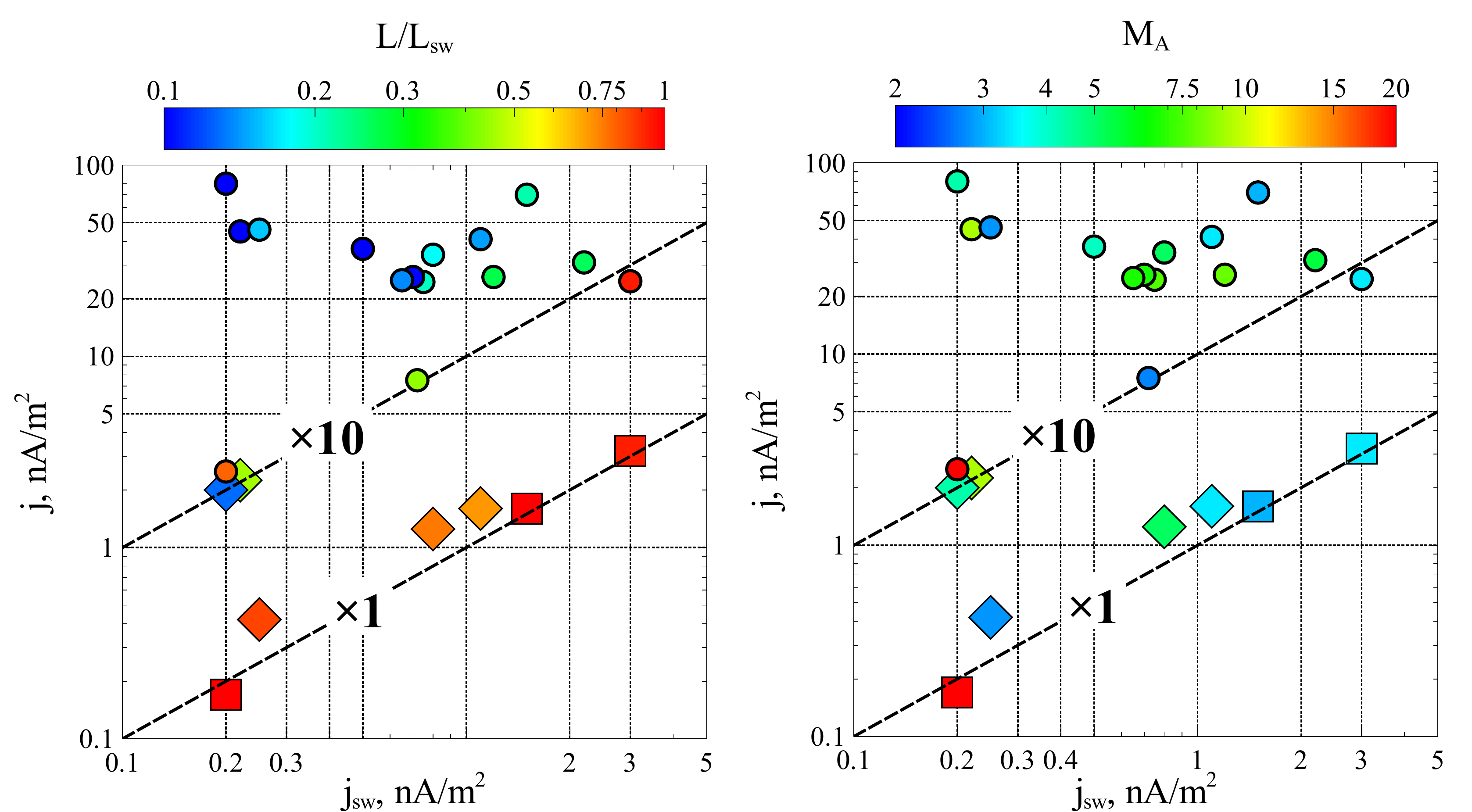}
\caption{\label{fig9} Circles show peak current density of solar wind discontinuities and discontinuities in the magnetosheath, after the bow shock crossing. Color codes: the ratio of discontinuity thickness in the downstream and foreshock normalized to the solar wind discontinuity thickness (left) and the Alfv\'{e}nic Mach number (right). Diamonds and squares show current peaks right upstream of the bow shock for five cases with reflected ions and three cases without reflected ions. }
\end{figure}

In addition to the four events shown above, we collected 11 events with similar spacecraft configurations: ARTEMIS observed a solar wind discontinuity that then is seen by THEMIS in the magnetosheath. In all these events, THEMIS also probed the bow shock configuration, but often we do not have discontinuity observations right in the foreshock, i.e. we can only compare the discontinuity well before the bow shock and right after the bow shock crossing. Figure \ref{fig9} shows the results of such a comparison.

There is a clear increase of the current density for all discontinuities: current density peaks in the magnetosheath are more than $10$ times larger than the corresponding peaks in the solar wind (circles). Such an increase does not depend on Alfv\'{e}n Mach number (right panel) nor on the shock wave normal angle (not shown), but seems to be more pronounced for initially weak discontinuities. The ratio of the discontinuity thicknesses in the magnetosheath and solar wind does not go below $1/10$ (left panel). Moreover, expression (\ref{eq:L}) underestimates the total thickness in the case of fluctuating magnetic fields. Actually, if $L < 0.1L_{sw}$ magnetic field oscillations must be present, which is one of the governing factors of the discontinuity current amplification. In other words, only in the oscillatory case may the current ratio significantly exceed $10$.

The magnetic field compression at the bow shock may also produce the current amplification by about an order of magnitude. Peculiarly, such a compression is several times greater than the Rankine-Hugoniot prediction and is not reproduced in our simulations (see Sect.~\ref{sec:sim}). Although kinetic effects such as the presence of reflected particles in the shock precursor actually lead to the compression being greater than that given by a simple hydrodynamic estimate, the difference is far less drastic. However, the observed local shock jump conditions are in a rough agreement with simulations.

The third and the smallest factor contributing to the amplification is the discontinuity thinning, taking place in the foreshock. There are eight events where discontinuities were also observed at the foreshock region, right before the bow shock crossing: three discontinuities are not accompanied by increased fluxes of reflected (hot) ions (boxes), and in the other five discontinuities we observe trapping of reflected ions (diamonds). For these five events, there is a current density increase and thickness decrease relative to the solar wind discontinuities. Thus, we can confirm the effect shown in Figure~\ref{fig2}: reflected ions alter the discontinuity configuration even before the discontinuity crosses the bow shock. This effect is well-observed in hybrid simulations (see below).

\section{Simulations}
\label{sec:sim}
In this section, we present results of a series of numerical simulations aimed at understanding the current density amplification during the discontinuity-shock interaction. We did not intend to reproduce any concrete observation discussed above, but rather to study the RD evolution in several illustrative cases, which let us reveal some general features.

For the simulations reported in this paper, we used our hybrid code ``Maximus" to investigate the RD interaction with the bow shock. The code is fully three-dimensional, divergence-free, and second-order accurate in time and space (see \citep{Kropotina2019} for a detailed description). The numerical convergence is discussed in Appendix \ref{appa}.

\subsection{Methods}
As supercritical collisionless shocks are mostly governed by ions' dynamics, the bow shock can be adequately modeled within the well-known hybrid approach \citep {Lipatov, Matthews94, Winske96}. In this framework: (i) the spatial region is discretized into cells with constant electromagnetic field within each cell; (ii) ions are represented as moving `macroparticles`(patches of phase space distribution); (iii) electrons are considered as a neutralizing massless fluid; and (iv) the displacement current is neglected in the Ampere's law due to the non-relativity of simulated flows. This approach leads to an algebraic instead of a differential equation for the electric field (the so-called generalized Ohm's law) and simplifies the main numerical scheme. Using hybrid codes instead of fully kinetic particle-in-cell codes substantially reduces the computing resources because ions' spatial and temporal scales are much larger than the electrons'.

We applied a standard normalization: the unit length is the far upstream proton inertial length $d_{p,u}$ (which is naturally different from the local values $d_p$ used in events \#1-\#4), the unit time is the proton inverse gyrofrequency $\Omega^{-1}$, and the unit density is the far upstream density $\rho_0$ (e.g.,  the unit velocity is the far upstream Alfv\'{e}n velocity). The corresponding typical values in the solar wind (note that we use SI units) are: $d_{p,u} = \sqrt{m_p/\mu_0 n e^2} = 230 / \sqrt{n_6}$ km $\sim 30-70$ km and $\Omega^{-1} = m_p / eB \approx (10/B_{-9})$ s $\sim 2$ s, where $m_p$ and $e$ are the proton mass and charge, and $n_6$ and $B_{-9}$ are the far upstream number density in cm$^{-3}$ and magnetic field in nT. Hence, the electric current is measured in units of $j_0 = e n \: \Omega \: d_{p,u} = 370 (\sqrt{n_6}/B_{-9})$ nA m$^{-2} \sim 500$ nA m$^{-2}$. Below, we present all the results in these units.

We used a fully 3D Cartesian grid with 50 particles per cell. We chose a 3D configuration to avoid possible dimensionality artifacts, which show up in 1D runs, especially for quasi-perpendicular shocks. The shock was initialized by pushing a supersonic super-Alfv\'{e}nic flow against a conducting reflective wall at the $x = 0$ plane. The flow is constantly injected through the open boundary at a large positive $x$. The $y$ and $z$ boundaries are periodic. The incoming and reflected flows produce a streaming instability, which in turn leads to formation of a planar shock front moving in the positive $x$ direction. All the results will be given in a frame where the downstream is at rest.

In each run, an RD was embedded into the upstream flow far enough from the wall to let the shock form before the interaction. The interaction was simulated in the frame of reference where the transverse velocity does not change its absolute value at the discontinuity (i.e. in the discontinuity normal incidence frame, which obviously is not the same as the shock normal incidence frame). We restricted our study to the case where the shock and discontinuity normal vectors are coaligned. A more complicated geometry requires a realistic shock front shape together with much greater transverse sizes. The shock-RD initial configuration is uniquely defined by five dimensionless parameters: \begin{itemize}
    \item shock Mach number $M_A$ (we give it in the front rest frame);
    \item shock inclination angle $\theta$, which in our case coincides with the RD inclination angle;
    \item upstream plasma beta (the ratio of the thermal to magnetic pressure: $\beta = 2\mu_0 n k T / B^2$, which typically is between about 0.1 to several tens in the solar wind. Hereafter we assume ions $\beta$ to be equal to the electrons beta, i.e. the total upstream $\beta$ is twice as large);
    \item discontinuity rotational angle $\Delta \phi$;
    \item initial RD width $D$.
\end{itemize}
In all the runs, we set $D = 2.5 d_{p,u}$ at $t = 0$, but $D$ quickly grew to a relatively constant far upstream value $L$ (that depended on other simulation parameters).

In terms of these parameters, the initial velocity and magnetic field (in dimensionless units) are:
\begin{eqnarray*}
\Phi(x) &=& 0.5\Delta \phi \left(1 - \tanh(\left(x - x_0)/D\right)\right); \\
_x &=& V_{RD} - \cos\theta; \quad V_y = - \sin\theta \cos\left(\Phi(x)\right); \quad V_z = -\sin\theta \sin\left(\Phi(x)\right); \\
B_x &=& \cos\theta; \quad B_y = \sin\theta\cos\left(\Phi(x)\right); \quad B_z = \sin\theta\cos\left(\Phi(x)\right).
\end{eqnarray*}
where $x_0$ is the initial RD position, $V_{RD}$ is the (negative) RD velocity relative to the reflective wall, and $|V_x|$ is plasma Alfv\'{e}n Mach number in downstream frame, so that $M_a$ can be found by adding the simulated front velocity to $|V_x|$.

Table \ref{tableofruns} lists all these parameters together with used grid sizes and resolution. Altogether, we studied two Mach numbers, three inclinations, and two rotational angles.

\begin{table}[]
    \centering
    \begin{tabular}{|c|c|c|c|c|c|c| }
    \hline
         title & $M_A$\footnote{Alfv\'{e}n Mach number in the shock rest frame}  & $\theta$ \footnote{shock and RD inclination angle} & $\beta$ \footnote{far upstream thermal to magnetic pressure ratio of ions} & $\Delta \phi$ \footnote{rotational angle}& box sizes [cells] & cell size $[d_{p,u}] $\\
         \hline
         A  & 7.8 & $75^\circ$ & 1 & $240^\circ$  & $5000 \times 100 \times 100$& 0.33\\
         \hline
         B & 7.8 & $15^\circ$ & 1 & $240^\circ$ & $7500 \times 100 \times 100$& 0.33\\
         \hline
         C & 7.8 & $15^\circ$ & 1 & $180^\circ$  & $7500 \times 100 \times 100$& 0.33\\
         \hline
         D & 5 & $15^\circ$ & 0.4 & $180^\circ$  & $7500 \times 50 \times 50$ & 0.5 \\
         \hline
         E & 5 & $40^\circ$ & 0.4 & $180^\circ$  & $6000 \times 100 \times 100$ & 0.33 \\
         \hline
    \end{tabular}
    \caption{Table of runs. \label{tableofruns}}
\end{table}

\subsection{Results for a quasi-perpendicular shock}

\begin{figure}[h]
\includegraphics[width=\linewidth]{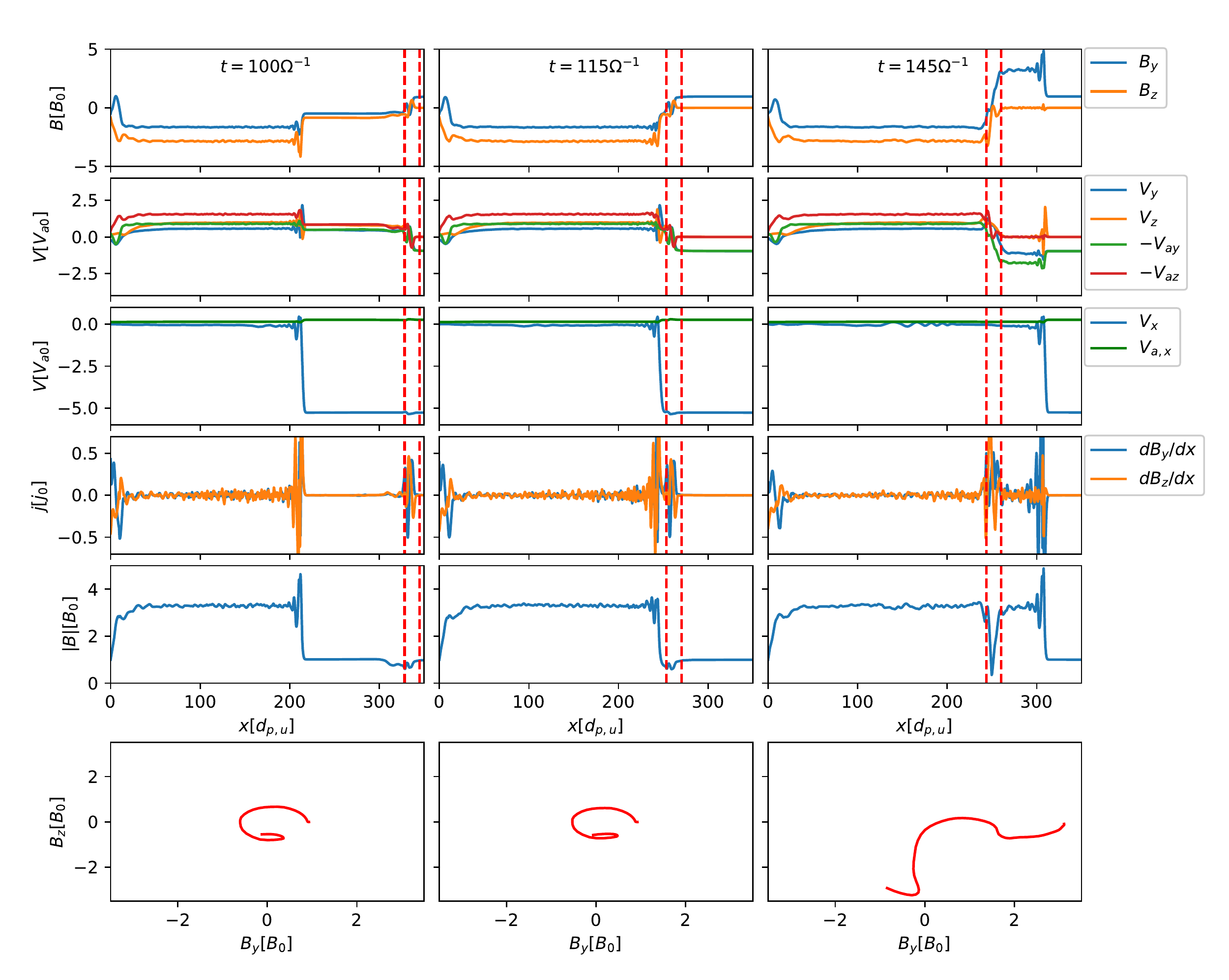}
\caption{An RD-shock interaction in three subsequent times for run A. From top to bottom: transverse magnetic field, transverse flow velocity and transverse Alfv\'{e}n velocity, parallel flow velocity and parallel Alfv\'{e}n velocity, current density, $|{\bf B}|$, and $B_y-B_z$ hodograms of the marked region near RD.}
\label{3timesperp_Ma7_8}
\end{figure}

The only quasi-perpendicular case in our simulations is run A. This is a relatively strong shock interacting with a strong discontinuity. In a simplified case of coinciding shock and RD normals, the shock upstream conditions uniquely determine the RD amplitude, which is equal to $B_0 \sin \theta$. This amplitude is quite large for all observational quasi-perpendicular shocks, except when the Alfv\'{e}n Mach number is high (see example event \#2).

Figure \ref{3timesperp_Ma7_8} illustrates three stages of the RD-shock interaction. The left column shows an isolated RD far upstream from the shock (in the pristine solar wind), the middle column shows the RD just upstream from the shock, and the right column shows the RD after the shock crossing. All spatial profiles are averaged over a cross section $x = {\rm const}$. The RD is boxed by the red dashed lines, and the hodograms in the bottom panels are drawn for the regions between these borders.

The supercritical quasi-perpendicular shock demonstrates a standard foot-ramp-overshoot structure. Such shocks are governed by magnetic mirror reflection of some of the incoming ions, but a strong transverse magnetic field prevents the reflected population from penetrating further upstream. Thus, they can neither form the shock precursor nor be accelerated  by  the first-order Fermi mechanism, and the whole upstream region remains unperturbed. The proton phase spaces reveal the absence of the upstream nonthermal population and electromagnetic waves, hence no shock-RD interaction is expected until the shock ramp. The downstream region is also calm, with nearly constant fields and flows.

Soon after the initiation, the discontinuity broadens from $D = 2.5d_{p,u}$ to a relatively stable width $L_{sw} \sim 17 d_{p,u}$ and then keeps its structure all the way to the shock, as expected. The Alfv\'{e}nic nature of the discontinuity can be checked by comparing Alfv\'{e}n and flow transverse velocities (see the second row in Figure~\ref{3timesperp_Ma7_8}). After the shock crossing, the transverse magnetic field increases by about $3.3$ times, but the current density remains nearly the same. The discontinuity becomes quite compressional, i.e. the pressure pulse is very strong.

The evolution of the magnetic field and current is shown in the top row in Figure~\ref{comparetransitions}. The simulated shock has $M_A = 7.8$, which is between $M_A = 20$ (event \#2, Figure~\ref{fig4}) and $M_A = 3$ (event \#3, Figure \ref{fig6}), and it clearly combines some properties of both observations.

One of key successes of the simulations is that they resolved a severe postshock $|{\bf B}|$ drop, also observed in observations (see the right panel in Figure \ref{fig6}), for a strong RD crossing a low Mach number shock. Both for observations and simulations, magnetic field profiles are similar in the solar wind and in the foreshock (i.e. negligible RD-shock interaction before approaching the shock ramp). This is natural for shocks without a precursor.

RD magnetic profiles after shock crossing are different in events \#2 and \#3. In the case of strong shock and weak RD, the latter keeps its shape while crossing the front (see the top row of Figure~\ref{fig4}). This means that, in units of local $d_p$, the profile broadens (the bottom row of the same figure). The same broadening is seen in our simulations (Figures~\ref{3timesperp_Ma7_8} and \ref{comparetransitions}). At the same time, a strong RD is thinned after having crossed a weak quasi-perpendicular shock (see Figure~\ref{fig6}).

Qualitatively, the thickness of any collisionless structure is determined by two competitive processes: the nonlinear steepening and the dispersive broadening \citep{Sagdeev1966, Treumann09, Balogh, Kennel85}. Thus, the same discontinuity can be thinned or broadened depending on which process wins. It is natural to suppose that high turbulence at a strong shock front can lead to some dispersive effects (e.g., high postshock magnetic fluctuations in panel (e) in Figure \ref{fig3}). To verify this, we perform a separate simulation (not shown) of the RD interaction with an Alfv\'{e}n wave; this simulation shows such interaction indeed smooths RD magnetic field profiles.

The opposite case of a weaker shock with a relatively smooth front (Figure \ref{fig5}) seems to be preferential for the discontinuity steepening. The reason is that the upstream flow velocity is decelerated gradually during shock crossing. Hence, the leading edge of the discontinuity always moves through the shock a bit slower than the trailing edge, which naturally causes the discontinuity thinning. This effect is especially important in the quasi-parallel shock precursor and will be discussed in the next subsection. Also note that there were some observational hints regarding oblique discontinuity orientation relative to the shock normal. This makes a shock crossing effectively longer and smoother.

In simulations, the current density remained almost unchanged during the shock transition. This is not the case in both observations, because of strong magnetic field amplification well above the Rankine-Hugoniot predictions. However, this effect cannot be obtained without taking into account the bow shock geometry and the Earth magnetic dipole. In all our simulations, $B_l$ compression does not exceed several units and the current density amplification is moderate.

\subsection{Results for quasi-parallel shocks}
\begin{figure}[h]
\includegraphics[width=\linewidth]{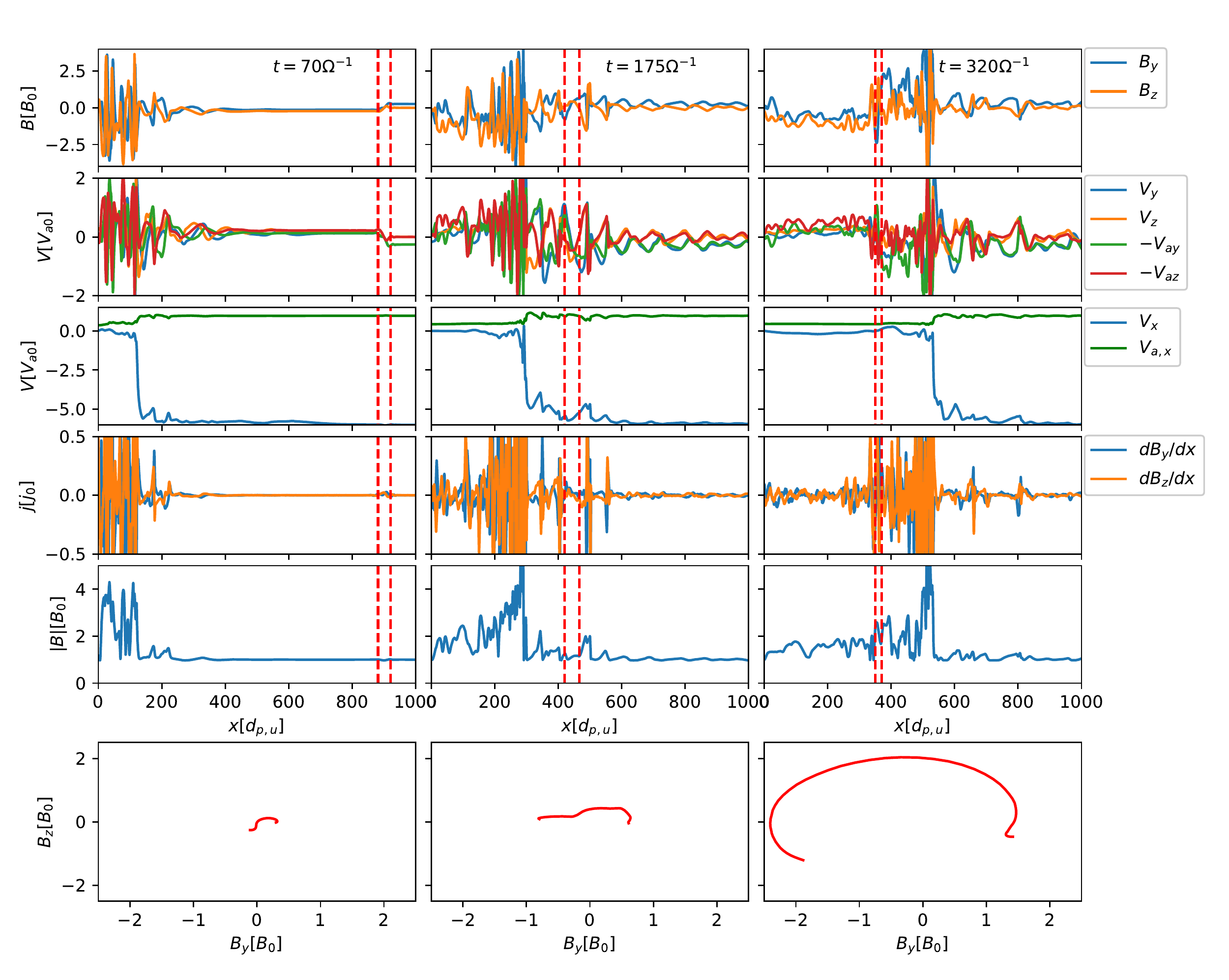}
\caption{The RD-shock interaction in three subsequent times for run B. From top to bottom: transverse magnetic field, transverse flow velocity and transverse Alfv\'{e}n velocity, parallel flow velocity and parallel Alfv\'{e}n velocity, current density, magnitude $|{\bf B}|$,  $B_y-B_z$ hodograms of the selected region near RD.} \label{3timespar}
\end{figure}

Besides the run discussed above, we also simulated the RD-shock configuration differing only by the inclination angle (see run B in Table \ref{tableofruns}, Figure \ref{3timespar} and the second row in Figure \ref{comparetransitions}). Because of a smaller inclination, the RD is initially weaker than in run A. At the same time, the shock is strong, and an effective turbulent RD broadening is expected. Moreover, the quasi-parallel shock precursor is occupied by nonthermal particles, which are nearly isotropic in the shock rest frame and produce a resonant streaming instability interacting with the upstream flow. Thus, an RD is substantially distorted by accelerated particles and plasma waves far before crossing the front. A train of shock-like secondary structures appears in the RD upstream (i.e. to the right from the RD position marked by red dashed lines). Simultaneously, moderately accelerated ({\it suprathermal}) particles are in turn affected by the RD: they hardly penetrate through the discontinuity, mostly remaining downstream of it and finally being swept downstream from the shock. The same effect was seen in all our quasi-parallel runs.

We simulated the RD interaction with a streaming suprathermal particles in the absence of a shock, and found the same effect. Interaction of accelerated particles with an RD caused its broadening and generated a train of secondary sharp structures, which in turn effectively trapped the accelerated particles. \textbf {Trapping of accelerated particles and possible related instabilities are discussed in Appendix B.}

Despite these kinetic effects, the RD hodogram remains mostly Alfv\'{e}nic-like (note that the $V_\perp$ and $-V_{a,\perp}$ profiles in the second row in Figure~\ref{3timespar} are very similar). The electric current grows from $j_{sw} \sim 0.03 j_0$ far upstream to $j_f \sim 0.11 j_0$ in the foreshock, partially because of $B_l$ doubling and partially due to plasma waves superimposed on the RD profile. The magnetosheath current reaches about $j_m \sim 0.8 j_0 \sim 27 j_{sw}$. Such an effective amplification is mainly caused by a substantial RD steepening as shown in Figure \ref{comparetransitions}. The discontinuity is clearly thinned even in terms of a local $d_p$, and in the absolute units the steepening is even more drastic ($d_p$ is reduced by almost a half downstream). On the other hand, at $t = 175 \Omega^{-1}$ the discontinuity thickness $L$ is comparable to its initial value, hence the steepening occurs later.

One of the reasons for such discontinuity modification is velocity gradient. Near the shock front, the upstream flow is decelerated by the pressure of accelerated particles and $V_{a,x}$ changes only slightly (see the third row in Figure \ref{3timespar}). An ideal MHD discontinuity's limits move with a local speed of $V_x + V_{a,x}$ \citep{LANDAU1984225}, so that the trailing edge of a discontinuity gradually overtakes the leading edge. For the simulation run with $M_A = 5$, we indeed observed that the discontinuity edges move with the local flow plus the Alfv\'{e}n velocity until they cross the front. For a higher $M_A = 8$, the RD was additionally slowed down in the immediate upstream.

The right column in Figure \ref{3timespar} shows that the discontinuity separates the downstream region occupied by strong magnetic field fluctuations from a relatively quiet region, which is in a good agreement with observations (event \#1). These oscillations partly interfere with the RD and lead to local changes of the direction of the current.

\subsection{Overview and discussion}
\begin{figure}[ht]
    \centering
    \includegraphics{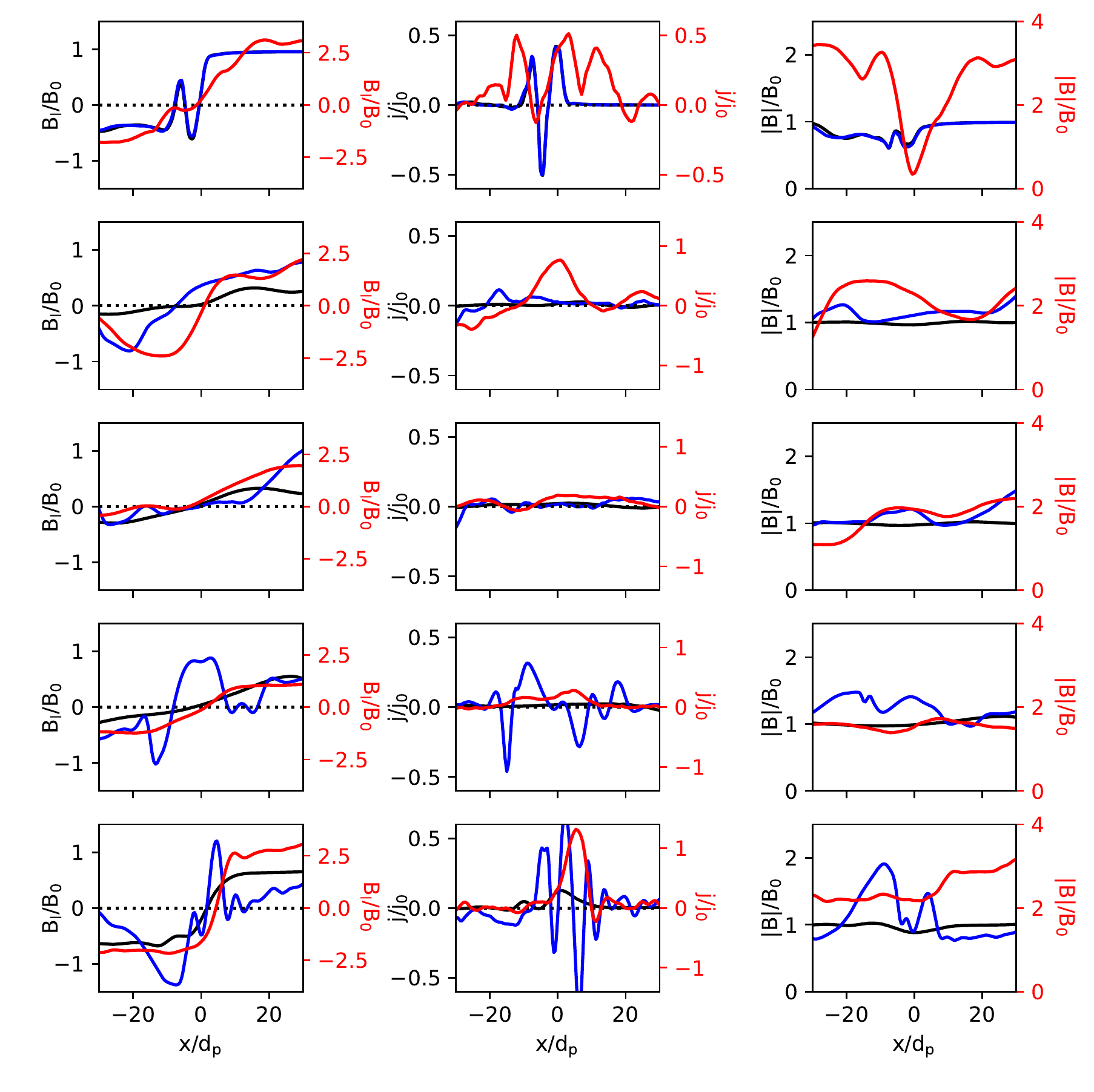}
    \caption{Magnetic field $B_l$, current density, and magnitude $|{\bf B}|$ for discontinuities from all simulations (A-E from top to bottom) in the solar wind (black), foreshock (blue), and magnetosheath (red). The spatial scale is normalized by the local proton inertial length.}
    \label{comparetransitions}
\end{figure}

Figure \ref{comparetransitions} illustrates the evolution of an RD during a shock crossing for all simulated cases, and Table \ref{tableofresults} lists current density and transverse magnetic field amplification factors. The key difference between quasi-parallel (runs B-D) and quasi-perpendicular (run A) shocks is that, in the latter case, a discontinuity does not change in the foreshock. At the same time, the extended quasi-parallel shock precursor modulates the RD and leads to the current density preamplification. Current density amplification in quasi-parallel shocks tends to be about $10$ times stronger and several times larger than the magnetic field compression. The latter is also slightly higher than the Hugoniot prediction (a well-known effect of accelerated particles pressure in the precursor \citep{Ellison90,Bykov14}).

The observations did not show a clear dependence of the current amplification on the inclination angle. The reason is that, in reality, energetic particles accelerated at a quasi-parallel region of the bow shock can penetrate into a quasi-perpendicular region and produce plasma instabilities, pressure gradients, and other effects capable of altering the discontinuity. Moreover, the accelerated particles can modify a quasi-perpendicular shock and be reaccelerated on it \citep{Caprioli2018}. Therefore, it is more convenient to distinguish shocks not by the inclination angle, but rather by the presence of a shock precursor filled with accelerated particles.

Figure \ref{fig9} illustrates the absence of preamplification in shocks without reflected ions. Two of three such events show the smallest current amplification in the whole dataset ($\sim 10$) and are not thinned during the shock crossing. The third event is example event \#3 with a still-moderate amplification factor $\sim 40$. In this event, the RD is substantially thinned during the shock crossing, which could be a consequence of a rather low, nearly subcritical, Mach number, which caused a smoother shock transition. In this case, the dispersive broadening is likely to be less effective than the nonlinear steepening.

Three cases with the strongest current amplification (the circles in the top left corner of Figure \ref{fig9}) correspond to the shock with reflected particles upstream (see diamonds of the same color in the right panel in Figure \ref{fig9}). Interestingly, all these cases involve weak discontinuities with small currents. However, this may be a case of survivorship bias: strong discontinuities after strong amplification are likely to reconnect very soon and hence cannot be observed in the downstream region.

And here are some additional observations:
\begin{itemize}
    \item Electric currents are substantially more amplified for the RDs with greater rotational angles (compare runs B and C in Table~\ref{tableofresults})
    \item There are no significant differences in current amplification between $M_A = 5$ and $M_A = 7.8$ (runs C and D), as well as between $\theta = 15^\circ$ and $\theta = 40^\circ$ (runs D and E). This is in a general agreement with observations.
    \item Quasi-perpendicular shocks are more likely to form strong pressure pulses, especially for discontinuities with high $\Delta\phi$ (see the right column in Figure~\ref{comparetransitions}).
\end{itemize}
However, these statements need to be verified against a larger set of observational and simulation statistics.

\begin{table}[]
    \centering
    \begin{tabular}{|c|c|c|c|c|c|c|c|c|}
    \hline
         Run & $j_{sw}$\footnote{current in the solar wind}  & $j_f$ \footnote{current in the foreshock} & $j_f / j_{sw}$ &$B_{l,f}/B_{l,sw}$ \footnote{transverse B compression in the foreshock} & $j_m$ \footnote{current in the magnetosheath} & $j_m/j_{sw}$ &$B_{l,m}/B_{l,sw}$ \footnote{transverse B compression in the magnetosheath}\\
         \hline
         A  & 0.4 & 0.4 & 1 &1 & 0.5 & 1.25 &3.5\\
         \hline
         B & 0.03 & 0.11 & 4 & 2 & 0.8 & 27 & 5.5\\
         \hline
         C & 0.025 & 0.07 & 2.8 & 2 & 0.2 & 8 & 5.5\\
         \hline
         D&  0.03 & 0.3 & 2.5 &10 & 0.3 & 10 & 4\\
         \hline
         E& 0.11 & 0.6 & 5 & 1.5 & 1.3 & 12 & 4.7\\
         \hline
    \end{tabular}
    \caption{RD current amplification during shock crossing} \label{tableofresults}
\end{table}

\section{Conclusions}
\label{sec:concl}
In this study, we investigated the bow shock crossing by an individual RD and showed that such crossings often result in the discontinuity thinning and current amplification. This is important because thinner (more intense) discontinuities are expected to be unstable to magnetic reconnections, which are often observed downstream of the Earth's bow shock and may contribute to the magnetosheath plasma heating \citep{Retino07:nat, Phan18:nature, Gingell20, Wang20:discontinuity, Bessho20}. Moreover, thinner discontinuities with a stronger current density are more effective in perturbing the Earth's magnetosphere boundary, the magnetopause \citep{Hietala18:jet, Chen21}. Therefore, our results indicate an important role of the discontinuity/bow shock interaction in the subsequent discontinuity impact on the space weather \citep[e.g.,][]{Kokubun77, Tsurutani11:jastp, Korotova12,Turc15}.

We used {\it in situ} observations and kinetic hybrid simulations to investigate the interaction of RDs with shocks described by different geometries. We found that the shock precursor occupied by energetic particles and plasma instabilities plays a crucial role in the current amplification. The amplification consists of three factors: transverse magnetic field compression, hydrodynamic discontinuity thinning in gradual parallel velocity profile, and superimposed magnetic field oscillations.

The first factor should provide a fourfold amplification due to the Hugoniot prediction on the magnetic field compression, but for the observed bow shock crossings this factor often reaches $10$. This effect cannot be explained within the model of an isolated planar shock. The second factor is actually rather weak, being balanced by the dispersive RD broadening. The third factor, which actually mimics the presence of postshock magnetic turbulence (or some small-scale waves of the external nature), seems to be crucial for the RD current amplification. This factor results in strong alternating current oscillations, which can shape very thin intense current layers and likely trigger magnetic reconnection.

Both observations and simulations suggest that the current amplification does not strongly depend on the shock Mach number. According to simulation results, the discontinuity/shock interaction is significantly different for quasi-perpendicular and quasi-parallel shock configurations. However, this difference is not so evident in observations. Our analysis indicates that the key property is not the inclination angle itself, but the presence of the shock precursor occupied by magnetic waves and accelerated particles. While in simulations such a precursor can appear only in quasi-parallel cases, in the real Earth's bow shock these accelerated particles can come from other bow shock regions (in particular, they can move along the discontinuity itself, if it is connected to the bow shock). Thus, quasi-perpendicular shocks can be very efficient current amplifiers as well.

Both observations and simulations indicate that different RDs can be amplified differently while crossing the same shock. Simulations suggest that the current amplification is more efficient for discontinuities with greater rotational angles.

\section*{Acknowledgements}
The authors thank the anonymous referee for a careful
reading of our paper and very constructive comments.
We acknowledge NASA contract NAS5-02099 for use of ARTEMIS data. We would like to thank C. W. Carlson and J. P. McFadden for use of ESA data, D.E. Larson and R.P. Lin for use of SST data, and K. H. Glassmeier, U. Auster, and W. Baumjohann for the use of FGM data provided under the lead of the Technical University of Braunschweig and with a financial support through the German Ministry for Economy and Technology and the German Aerospace Center (DLR) under contract 50 OC 0302. ARTEMIS data were downloaded from http://themis.ssl.berkeley.edu/. Data access and processing was done using SPEDAS V3.1, see \citet{Angelopoulos19}. J.A.K. and A.M.B. were supported by the RSF grant 21-72-20020. Some of the modeling was performed at the Joint Supercomputer Center JSCC RAS and at the ``Tornado'' subsystem of the St.~Petersburg Polytechnic University supercomputing center. A.V.A. was supported by NASA grant \#80NSSC20K1788. I.Y.V. was supported by NASA grant \#80NSSC20K1781.


\begin{thebibliography}{}
\expandafter\ifx\csname natexlab\endcsname\relax\def\natexlab#1{#1}\fi
\providecommand{\url}[1]{\href{#1}{#1}}
\providecommand{\dodoi}[1]{doi:~\href{http://doi.org/#1}{\nolinkurl{#1}}}
\providecommand{\doeprint}[1]{\href{http://ascl.net/#1}{\nolinkurl{http://ascl.net/#1}}}
\providecommand{\doarXiv}[1]{\href{https://arxiv.org/abs/#1}{\nolinkurl{https://arxiv.org/abs/#1}}}

\bibitem[{{An} {et~al.}(2020){An}, {Liu}, {Bortnik}, {Osmane}, \&
  {Angelopoulos}}]{An20:apj}
{An}, X., {Liu}, T.~Z., {Bortnik}, J., {Osmane}, A., \& {Angelopoulos}, V.
  2020, \apj, 901, 73, \dodoi{10.3847/1538-4357/abaf03}

\bibitem[{{Angelopoulos}(2008)}]{Angelopoulos08:ssr}
{Angelopoulos}, V. 2008, \ssr, 141, 5, \dodoi{10.1007/s11214-008-9336-1}

\bibitem[{{Angelopoulos}(2011)}]{Angelopoulos11:ARTEMIS}
---. 2011, \ssr, 165, 3, \dodoi{10.1007/s11214-010-9687-2}

\bibitem[{{Angelopoulos} {et~al.}(2019){Angelopoulos}, {Cruce}, {Drozdov},
  {Grimes}, {Hatzigeorgiu}, {King}, {Larson}, {Lewis}, {McTiernan}, {Roberts},
  {Russell}, {Hori}, {Kasahara}, {Kumamoto}, {Matsuoka}, {Miyashita},
  {Miyoshi}, {Shinohara}, {Teramoto}, {Faden}, {Halford}, {McCarthy}, {Millan},
  {Sample}, {Smith}, {Woodger}, {Masson}, {Narock}, {Asamura}, {Chang},
  {Chiang}, {Kazama}, {Keika}, {Matsuda}, {Segawa}, {Seki}, {Shoji}, {Tam},
  {Umemura}, {Wang}, {Wang}, {Redmon}, {Rodriguez}, {Singer}, {Vandegriff},
  {Abe}, {Nose}, {Shinbori}, {Tanaka}, {UeNo}, {Andersson}, {Dunn}, {Fowler},
  {Halekas}, {Hara}, {Harada}, {Lee}, {Lillis}, {Mitchell}, {Argall},
  {Bromund}, {Burch}, {Cohen}, {Galloy}, {Giles}, {Jaynes}, {Le Contel}, {Oka},
  {Phan}, {Walsh}, {Westlake}, {Wilder}, {Bale}, {Livi}, {Pulupa},
  {Whittlesey}, {DeWolfe}, {Harter}, {Lucas}, {Auster}, {Bonnell}, {Cully},
  {Donovan}, {Ergun}, {Frey}, {Jackel}, {Keiling}, {Korth}, {McFadden},
  {Nishimura}, {Plaschke}, {Robert}, {Turner}, {Weygand}, {Candey}, {Johnson},
  {Kovalick}, {Liu}, {McGuire}, {Breneman}, {Kersten}, \&
  {Schroeder}}]{Angelopoulos19}
{Angelopoulos}, V., {Cruce}, P., {Drozdov}, A., {et~al.} 2019, \ssr, 215, 9,
  \dodoi{10.1007/s11214-018-0576-4}

\bibitem[{{Archer} {et~al.}(2012){Archer}, {Horbury}, \& {Eastwood}}]{Archer12}
{Archer}, M.~O., {Horbury}, T.~S., \& {Eastwood}, J.~P. 2012, Journal of
  Geophysical Research, 117, A05228, \dodoi{10.1029/2011JA017468}

\bibitem[{{Arons}(2012)}]{Arons12}
{Arons}, J. 2012, \ssr, 173, 341, \dodoi{10.1007/s11214-012-9885-1}

\bibitem[{{Artemyev} {et~al.}(2018{\natexlab{a}}){Artemyev}, {Angelopoulos},
  {Halekas}, {Vinogradov}, {Vasko}, \& {Zelenyi}}]{Artemyev18:apj}
{Artemyev}, A.~V., {Angelopoulos}, V., {Halekas}, J.~S., {et~al.}
  2018{\natexlab{a}}, \apj, 859, 95, \dodoi{10.3847/1538-4357/aabe89}

\bibitem[{{Artemyev} {et~al.}(2018{\natexlab{b}}){Artemyev}, {Angelopoulos}, \&
  {McTiernan}}]{Artemyev18:jgr:report}
{Artemyev}, A.~V., {Angelopoulos}, V., \& {McTiernan}, J.~M.
  2018{\natexlab{b}}, Journal of Geophysical Research (Space Physics), 123,
  9955, \dodoi{10.1029/2018JA025904}

\bibitem[{{Artemyev} {et~al.}(2019{\natexlab{a}}){Artemyev}, {Angelopoulos}, \&
  {Vasko}}]{Artemyev19:jgr:solarwind}
{Artemyev}, A.~V., {Angelopoulos}, V., \& {Vasko}, I.~Y. 2019{\natexlab{a}},
  Journal of Geophysical Research (Space Physics), 124, 3858,
  \dodoi{10.1029/2019JA026597}

\bibitem[{{Artemyev} {et~al.}(2019{\natexlab{b}}){Artemyev}, {Angelopoulos},
  {Vasko}, {Runov}, {Avanov}, {Giles}, {Russell}, \&
  {Strangeway}}]{Artemyev19:grl:solarwind}
{Artemyev}, A.~V., {Angelopoulos}, V., {Vasko}, I.-Y., {et~al.}
  2019{\natexlab{b}}, \grl, 46, 1185–1194, \dodoi{10.1029/2018GL079906}

\bibitem[{{Auster} {et~al.}(2008){Auster}, {Glassmeier}, {Magnes}, {Aydogar},
  {Baumjohann}, {Constantinescu}, {Fischer}, {Fornacon}, {Georgescu}, {Harvey},
  {Hillenmaier}, {Kroth}, {Ludlam}, {Narita}, {Nakamura}, {Okrafka},
  {Plaschke}, {Richter}, {Schwarzl}, {Stoll}, {Valavanoglou}, \&
  {Wiedemann}}]{Auster08:THEMIS}
{Auster}, H.~U., {Glassmeier}, K.~H., {Magnes}, W., {et~al.} 2008, \ssr, 141,
  235, \dodoi{10.1007/s11214-008-9365-9}

\bibitem[{Balogh \& Treumann(2013)}]{Balogh}
Balogh, A., \& Treumann, R.~A. 2013, {Physics of Collisionless Shocks: Space
  Plasma Shock Waves}, ISSI Scientific Report Series (New York: Springer),
  \dodoi{10.1007/978-1-4614-6099-2}

\bibitem[{{Bessho} {et~al.}(2020){Bessho}, {Chen}, {Wang}, {Hesse}, {Wilson},
  \& {Ng}}]{Bessho20}
{Bessho}, N., {Chen}, L.~J., {Wang}, S., {et~al.} 2020, Physics of Plasmas, 27,
  092901, \dodoi{10.1063/5.0012443}

\bibitem[{{Bogovalov}(1999)}]{Bogovalov99}
{Bogovalov}, S.~V. 1999, Astronomy and Astrophysics, 349, 1017

\bibitem[{{Borovsky}(2010)}]{Borovsky10:solarwind}
{Borovsky}, J.~E. 2010, Physical Review Letters, 105, 111102,
  \dodoi{10.1103/PhysRevLett.105.111102}

\bibitem[{{Buneman}(1993)}]{Buneman1993}
{Buneman}, O. 1993, in Computer Space Plasma Physics: Simulation Techniques and
  Software, ed. H.~Matsumoto \& Y.~Omura (Tokyo: Terra Scientific Publishing
  Company (TERRAPUB)), 67--84

\bibitem[{{Bykov} {et~al.}(2017{\natexlab{a}}){Bykov}, {Amato}, {Petrov},
  {Krassilchtchikov}, \& {Levenfish}}]{AB17}
{Bykov}, A.~M., {Amato}, E., {Petrov}, A.~E., {Krassilchtchikov}, A.~M., \&
  {Levenfish}, K.~P. 2017{\natexlab{a}}, \ssr, 207, 235,
  \dodoi{10.1007/s11214-017-0371-7}

\bibitem[{{Bykov} {et~al.}(2017{\natexlab{b}}){Bykov}, {Ellison}, \&
  {Osipov}}]{Bykov2017}
{Bykov}, A.~M., {Ellison}, D.~C., \& {Osipov}, S.~M. 2017{\natexlab{b}}, \pre,
  95, 033207, \dodoi{10.1103/PhysRevE.95.033207}

\bibitem[{{Bykov} {et~al.}(2014){Bykov}, {Ellison}, {Osipov}, \&
  {Vladimirov}}]{Bykov14}
{Bykov}, A.~M., {Ellison}, D.~C., {Osipov}, S.~M., \& {Vladimirov}, A.~E. 2014,
  \apj, 789, 137, \dodoi{10.1088/0004-637X/789/2/137}

\bibitem[{{Bykov} {et~al.}(2013){Bykov}, {Gladilin}, \& {Osipov}}]{Bykov13}
{Bykov}, A.~M., {Gladilin}, P.~E., \& {Osipov}, S.~M. 2013, \mnras, 429, 2755,
  \dodoi{10.1093/mnras/sts553}

\bibitem[{{Bykov} {et~al.}(2019{\natexlab{a}}){Bykov}, {Petrov},
  {Krassilchtchikov}, {Levenfish}, {Osipov}, \& {Pavlov}}]{AB19}
{Bykov}, A.~M., {Petrov}, A.~E., {Krassilchtchikov}, A.~M., {et~al.}
  2019{\natexlab{a}}, \apjl, 876, L8, \dodoi{10.3847/2041-8213/ab1922}

\bibitem[{{Bykov} {et~al.}(2019{\natexlab{b}}){Bykov}, {Vazza}, {Kropotina},
  {Levenfish}, \& {Paerels}}]{Bykov2019}
{Bykov}, A.~M., {Vazza}, F., {Kropotina}, J.~A., {Levenfish}, K.~P., \&
  {Paerels}, F.~B.~S. 2019{\natexlab{b}}, \ssr, 215, 14,
  \dodoi{10.1007/s11214-019-0585-y}

\bibitem[{{Cable} \& {Lin}(1998)}]{Cable&Lin98}
{Cable}, S., \& {Lin}, Y. 1998, \jgr, 103, 29551, \dodoi{10.1029/1998JA900025}

\bibitem[{{Caprioli} {et~al.}(2018){Caprioli}, {Zhang}, \&
  {Spitkovsky}}]{Caprioli2018}
{Caprioli}, D., {Zhang}, H., \& {Spitkovsky}, A. 2018, Journal of Plasma
  Physics, 84, 715840301, \dodoi{10.1017/S0022377818000478}

\bibitem[{{Chen} {et~al.}(2021){Chen}, {Ng}, {Omelchenko}, \& {Wang}}]{Chen21}
{Chen}, L.-J., {Ng}, J., {Omelchenko}, Y., \& {Wang}, S. 2021, arXiv e-prints,
  arXiv:2103.07448.
\newblock \doarXiv{2103.07448}

\bibitem[{{Cornwall}(1965)}]{Cornwall1965}
{Cornwall}, J.~M. 1965, \jgr, 70, 61, \dodoi{10.1029/JZ070i001p00061}

\bibitem[{{Davis} {et~al.}(2020){Davis}, {Cattell}, {Wilson}, {Cohen},
  {Breneman}, \& {Hanson}}]{Davis20}
{Davis}, L., {Cattell}, C.~A., {Wilson}, L.~B., I., {et~al.} 2020, arXiv
  e-prints, arXiv:2006.01064.
\newblock \doarXiv{2006.01064}

\bibitem[{{de Keyser} {et~al.}(1996){de Keyser}, {Roth}, {Lemaire},
  {Tsurutani}, {Ho}, \& {Hammond}}]{deKeyser96}
{de Keyser}, J., {Roth}, M., {Lemaire}, J., {et~al.} 1996, \solphys, 166, 415,
  \dodoi{10.1007/BF00149407}

\bibitem[{{de Keyser} {et~al.}(1997){de Keyser}, {Roth}, {Tsurutani}, {Ho}, \&
  {Phillips}}]{deKeyser97}
{de Keyser}, J., {Roth}, M., {Tsurutani}, B.~T., {Ho}, C.~M., \& {Phillips},
  J.~L. 1997, \aap, 321, 945

\bibitem[{{Drake} {et~al.}(2020){Drake}, {Agapitov}, {Swisdak}, {Badman},
  {Bale}, {Horbury}, {Kasper}, {MacDowall}, {Mozer}, {Phan}, {Pulupa}, {Szabo},
  \& {Velli}}]{Drake20}
{Drake}, J.~F., {Agapitov}, O., {Swisdak}, M., {et~al.} 2020, arXiv e-prints,
  arXiv:2009.05645.
\newblock \doarXiv{2009.05645}

\bibitem[{{Ellison} {et~al.}(1990){Ellison}, {Moebius}, \&
  {Paschmann}}]{Ellison90}
{Ellison}, D.~C., {Moebius}, E., \& {Paschmann}, G. 1990, \apj, 352, 376,
  \dodoi{10.1086/168544}

\bibitem[{{Farrugia} {et~al.}(1995){Farrugia}, {Erkaev}, {Biernat}, \&
  {Burlaga}}]{Farrugia95}
{Farrugia}, C.~J., {Erkaev}, N.~V., {Biernat}, H.~K., \& {Burlaga}, L.~F. 1995,
  \jgr, 100, 19245, \dodoi{10.1029/95JA01080}

\bibitem[{{Farrugia} {et~al.}(2013){Farrugia}, {Erkaev}, {Jordanova}, {Lugaz},
  {Sandholt}, {M{\"u}hlbachler}, \& {Torbert}}]{Farrugia13}
{Farrugia}, C.~J., {Erkaev}, N.~V., {Jordanova}, V.~K., {et~al.} 2013, Journal
  of Atmospheric and Solar-Terrestrial Physics, 99, 14,
  \dodoi{10.1016/j.jastp.2012.11.014}

\bibitem[{{Farrugia} {et~al.}(2018){Farrugia}, {Cohen}, {Vasquez}, {Lugaz},
  {Alm}, {Torbert}, {Argall}, {Paulson}, {Lavraud}, {Gershman}, {Gratton},
  {Matsui}, {Rogers}, {Forbes}, {Payne}, {Ergun}, {Mauk}, {Burch}, {Russell},
  {Strangeway}, {Shuster}, {Nakamura}, {Fuselier}, {Giles}, {Khotyaintsev},
  {Lindqvist}, {Marklund}, {Petrinec}, \& {Pollock}}]{Farrugia18}
{Farrugia}, C.~J., {Cohen}, I.~J., {Vasquez}, B.~J., {et~al.} 2018, Journal of
  Geophysical Research (Space Physics), 123, 8983, \dodoi{10.1029/}

\bibitem[{{Galeev} \& {Sudan}(1985)}]{bookGaleev85:vol2}
{Galeev}, A.~A., \& {Sudan}, R.~N. 1985, {Handbook of plasma physics. Vol. 2:
  Basic plasma physics II.}

\bibitem[{Gary {et~al.}(2017)Gary, Fu, Cowee, Winske, \& Liu}]{Gary2017}
Gary, S.~P., Fu, X., Cowee, M.~M., Winske, D., \& Liu, K. 2017, Journal of
  Geophysical Research: Space Physics, 122, 464,
  \dodoi{https://doi.org/10.1002/2016JA023425}

\bibitem[{{Gedalin} {et~al.}(2020){Gedalin}, {Zhou}, {Russell}, \&
  {Angelopoulos}}]{Gedalin20:angeo}
{Gedalin}, M., {Zhou}, X., {Russell}, C.~T., \& {Angelopoulos}, V. 2020,
  Annales Geophysicae, 38, 17, \dodoi{10.5194/angeo-38-17-2020}

\bibitem[{{Gingell} {et~al.}(2020){Gingell}, {Schwartz}, {Eastwood}, {Stawarz},
  {Burch}, {Ergun}, {Fuselier}, {Gershman}, {Giles}, {Khotyaintsev}, {Lavraud},
  {Lindqvist}, {Paterson}, {Phan}, {Russell}, {Strangeway}, {Torbert}, \&
  {Wilder}}]{Gingell20}
{Gingell}, I., {Schwartz}, S.~J., {Eastwood}, J.~P., {et~al.} 2020, Journal of
  Geophysical Research (Space Physics), 125, e27119,
  \dodoi{10.1029/2019JA027119}

\bibitem[{{Goncharov} {et~al.}(2015){Goncharov}, {{\v{S}}afr{\'a}nkov{\'a}}, \&
  {N{\v{e}}me{\v{c}}ek}}]{Goncharov15}
{Goncharov}, O., {{\v{S}}afr{\'a}nkov{\'a}}, J., \& {N{\v{e}}me{\v{c}}ek}, Z.
  2015, \planss, 115, 4, \dodoi{10.1016/j.pss.2014.12.001}

\bibitem[{{Gosling}(2012)}]{Gosling12}
{Gosling}, J.~T. 2012, \ssr, 172, 187, \dodoi{10.1007/s11214-011-9747-2}

\bibitem[{{Greco} {et~al.}(2009{\natexlab{a}}){Greco}, {Matthaeus}, {Servidio},
  {Chuychai}, \& {Dmitruk}}]{Greco09}
{Greco}, A., {Matthaeus}, W.~H., {Servidio}, S., {Chuychai}, P., \& {Dmitruk},
  P. 2009{\natexlab{a}}, \apjl, 691, L111, \dodoi{10.1088/0004-637X/691/2/L111}

\bibitem[{{Greco} {et~al.}(2009{\natexlab{b}}){Greco}, {Matthaeus}, {Servidio},
  \& {Dmitruk}}]{Greco09:pre}
{Greco}, A., {Matthaeus}, W.~H., {Servidio}, S., \& {Dmitruk}, P.
  2009{\natexlab{b}}, \pre, 80, 046401, \dodoi{10.1103/PhysRevE.80.046401}

\bibitem[{{Greco} {et~al.}(2016){Greco}, {Perri}, {Servidio}, {Yordanova}, \&
  {Veltri}}]{Greco16}
{Greco}, A., {Perri}, S., {Servidio}, S., {Yordanova}, E., \& {Veltri}, P.
  2016, \apjl, 823, L39, \dodoi{10.3847/2041-8205/823/2/L39}

\bibitem[{{Guo} {et~al.}(2021){Guo}, {Lin}, \& {Wang}}]{Guo21}
{Guo}, Z., {Lin}, Y., \& {Wang}, X. 2021, Journal of Geophysical Research:
  Space Physics, e2020JA028853, \dodoi{10.1029/2020JA028853}

\bibitem[{{Hietala} {et~al.}(2018){Hietala}, {Phan}, {Angelopoulos},
  {Oieroset}, {Archer}, {Karlsson}, \& {Plaschke}}]{Hietala18:jet}
{Hietala}, H., {Phan}, T.~D., {Angelopoulos}, V., {et~al.} 2018, \grl, 45,
  1732, \dodoi{10.1002/2017GL076525}

\bibitem[{{Hietala} {et~al.}(2012){Hietala}, {Sandroos}, \&
  {Vainio}}]{Hietala12:apj}
{Hietala}, H., {Sandroos}, A., \& {Vainio}, R. 2012, \apjl, 751, L14,
  \dodoi{10.1088/2041-8205/751/1/L14}

\bibitem[{{Horbury} {et~al.}(2001){Horbury}, {Burgess}, {Fr{\"a}nz}, \&
  {Owen}}]{Horbury01}
{Horbury}, T.~S., {Burgess}, D., {Fr{\"a}nz}, M., \& {Owen}, C.~J. 2001, \grl,
  28, 677, \dodoi{10.1029/2000GL000121}

\bibitem[{{Hubert} \& {Harvey}(2000)}]{Hubert&Harvey00}
{Hubert}, D., \& {Harvey}, C.~C. 2000, \grl, 27, 3149,
  \dodoi{10.1029/2000GL003776}

\bibitem[{{Karimabadi} {et~al.}(2005){Karimabadi}, {Daughton}, \&
  {Quest}}]{KDQ05theory}
{Karimabadi}, H., {Daughton}, W., \& {Quest}, K.~B. 2005, J. Geophys. Res.,
  110, 3214, \dodoi{10.1029/2004JA010749}

\bibitem[{{Karimabadi} {et~al.}(1995){Karimabadi}, {Krauss-Varban}, \&
  {Omidi}}]{Karimabadi95}
{Karimabadi}, H., {Krauss-Varban}, D., \& {Omidi}, N. 1995, \grl, 22, 2989,
  \dodoi{10.1029/95GL02887}

\bibitem[{{Keika} {et~al.}(2009){Keika}, {Nakamura}, {Baumjohann},
  {Angelopoulos}, {Kabin}, {Glassmeier}, {Sibeck}, {Magnes}, {Auster},
  {Forna{\c c}on}, {McFadden}, {Carlson}, {Lucek}, {Carr}, {Dandouras}, \&
  {Rankin}}]{Keika09}
{Keika}, K., {Nakamura}, R., {Baumjohann}, W., {et~al.} 2009, \jgr, 114,
  A00C26, \dodoi{10.1029/2008JA013481}

\bibitem[{{Kennel} \& {Coroniti}(1984)}]{Kennel&Coroniti84}
{Kennel}, C.~F., \& {Coroniti}, F.~V. 1984, \apj, 283, 710,
  \dodoi{10.1086/162357}

\bibitem[{{Kennel} {et~al.}(1985){Kennel}, {Edmiston}, \& {Hada}}]{Kennel85}
{Kennel}, C.~F., {Edmiston}, J.~P., \& {Hada}, T. 1985, Washington DC American
  Geophysical Union Geophysical Monograph Series, 34, 1,
  \dodoi{10.1029/GM034p0001}

\bibitem[{{Knetter} {et~al.}(2004){Knetter}, {Neubauer}, {Horbury}, \&
  {Balogh}}]{Knetter04}
{Knetter}, T., {Neubauer}, F.~M., {Horbury}, T., \& {Balogh}, A. 2004, \jgr,
  109, A06102, \dodoi{10.1029/2003JA010099}

\bibitem[{{Kokubun} {et~al.}(1977){Kokubun}, {McPherron}, \&
  {Russell}}]{Kokubun77}
{Kokubun}, S., {McPherron}, R.~L., \& {Russell}, C.~T. 1977, \jgr, 82, 74,
  \dodoi{10.1029/JA082i001p00074}

\bibitem[{{Korotova} {et~al.}(2012){Korotova}, {Sibeck}, {Omidi}, \&
  {Angelopoulos}}]{Korotova12}
{Korotova}, G.~I., {Sibeck}, D.~G., {Omidi}, N., \& {Angelopoulos}, V. 2012,
  Journal of Geophysical Research (Space Physics), 117, A12207,
  \dodoi{10.1029/2012JA017510}

\bibitem[{{Koval} {et~al.}(2005){Koval}, {{\v{S}}afr{\'a}nkov{\'a}},
  {N{\v{e}}me{\v{c}}ek}, {P{\v{r}}ech}, {Samsonov}, \& {Richardson}}]{Koval05b}
{Koval}, A., {{\v{S}}afr{\'a}nkov{\'a}}, J., {N{\v{e}}me{\v{c}}ek}, Z.,
  {et~al.} 2005, \grl, 32, L15101, \dodoi{10.1029/2005GL023009}

\bibitem[{{Koval} {et~al.}(2006){Koval}, {{\v{S}}afr{\'a}nkov{\'a}},
  {N{\v{e}}me{\v{c}}ek}, {Samsonov}, {P{\v{r}}ech}, {Richardson}, \&
  {Hayosh}}]{Koval06}
---. 2006, \grl, 33, L11102, \dodoi{10.1029/2006GL025707}

\bibitem[{{Krasnoselskikh} {et~al.}(2020){Krasnoselskikh}, {Larosa},
  {Agapitov}, {de Wit}, {Moncuquet}, {Mozer}, {Stevens}, {Bale}, {Bonnell},
  {Froment}, {Goetz}, {Goodrich}, {Harvey}, {Kasper}, {MacDowall}, {Malaspina},
  {Pulupa}, {Raouafi}, {Revillet}, {Velli}, \& {Wygant}}]{Krasnoselskikh20}
{Krasnoselskikh}, V., {Larosa}, A., {Agapitov}, O., {et~al.} 2020, \apj, 893,
  93, \dodoi{10.3847/1538-4357/ab7f2d}

\bibitem[{{Kropotina} {et~al.}(2018{\natexlab{a}}){Kropotina}, {Bykov},
  {Krassilchtchikov}, \& {Levenfish}}]{Kropotina2018}
{Kropotina}, J.~A., {Bykov}, A.~M., {Krassilchtchikov}, A.~M., \& {Levenfish},
  K.~P. 2018{\natexlab{a}}, in Journal of Physics Conference Series, Vol. 1038,
  Journal of Physics Conference Series, 012014,
  \dodoi{10.1088/1742-6596/1038/1/012014}

\bibitem[{{Kropotina} {et~al.}(2018{\natexlab{b}}){Kropotina}, {Bykov},
  {Krassilchtchikov}, \& {Levenfish}}]{Kropotina2019}
{Kropotina}, J.~A., {Bykov}, A.~M., {Krassilchtchikov}, A.~M., \& {Levenfish},
  K.~P. 2018{\natexlab{b}}, arXiv e-prints, arXiv:1806.05926.
\newblock \doarXiv{1806.05926}

\bibitem[{{Kropotina} {et~al.}(2020){Kropotina}, {Bykov}, {Osipov}, {Ermolina},
  \& {Romansky}}]{Kropotina2020}
{Kropotina}, J.~A., {Bykov}, A.~M., {Osipov}, S.~M., {Ermolina}, V.~E., \&
  {Romansky}, V.~I. 2020, Journal of Technical Physics, 65, 14,
  \dodoi{10.1134/S1063784220010144}

\bibitem[{{Kropotina} {et~al.}(2019){Kropotina}, {Levenfish}, \&
  {Bykov}}]{Kropotina2019conf}
{Kropotina}, J.~A., {Levenfish}, K.~P., \& {Bykov}, A.~M. 2019, in Journal of
  Physics Conference Series, Vol. 1400, Journal of Physics Conference Series,
  022002, \dodoi{10.1088/1742-6596/1400/2/022002}

\bibitem[{{Kropotina} {et~al.}(2018{\natexlab{c}}){Kropotina}, {Bykov},
  {Kozlova}, {Krassilchtchikov}, {Levenfish}, \&
  {Blinnikov}}]{Kropotina2018PAN}
{Kropotina}, Y.~A., {Bykov}, A.~M., {Kozlova}, A.~V., {et~al.}
  2018{\natexlab{c}}, Physics of Atomic Nuclei, 81, 139,
  \dodoi{10.1134/S1063778818010155}

\bibitem[{{Kuznetsova} \& {Roth}(1995)}]{Kuznetsova95}
{Kuznetsova}, M.~M., \& {Roth}, M. 1995, J. Geophys. Res., 100, 155,
  \dodoi{10.1029/94JA02329}

\bibitem[{LANDAU \& LIFSHITZ(1984)}]{LANDAU1984225}
LANDAU, L., \& LIFSHITZ, E. 1984, in Course of Theoretical Physics, Vol.~8,
  Electrodynamics of Continuous Media (Second Edition), second edition edn.,
  ed. L.~LANDAU \& E.~LIFSHITZ (Amsterdam: Pergamon), 225--256,
  \dodoi{https://doi.org/10.1016/B978-0-08-030275-1.50014-X}

\bibitem[{{Larosa} {et~al.}(2020){Larosa}, {Krasnoselskikh}, {Dudok de
  Wit{\i}nst}, {Agapitov}, {Froment}, {Jagarlamudi}, {Velli}, {Bale}, {Case},
  {Goetz}, {Harvey}, {Kasper}, {Korreck}, {Larson}, {MacDowall}, {Malaspina},
  {Pulupa}, {Revillet}, \& {Stevens}}]{Larosa20:arXiv}
{Larosa}, A., {Krasnoselskikh}, V., {Dudok de Wit{\i}nst}, T., {et~al.} 2020,
  arXiv e-prints, arXiv:2012.10420.
\newblock \doarXiv{2012.10420}

\bibitem[{{Le Veque} {et~al.}(1998){Le Veque}, {Mihalas}, {Dorfi}, \&
  {Müller}}]{leveque}
{Le Veque}, R., {Mihalas}, D., {Dorfi}, E., \& {Müller}, E. 1998,
  Computational Methods for Astrophysical Fluid Flow (Berlin, Heidelberg:
  Springer)

\bibitem[{{Lin} {et~al.}(2009){Lin}, {Tsai}, {Chen}, {Weng}, {Chao}, \&
  {Lee}}]{Lin09:rotational_discontinuities}
{Lin}, C.~C., {Tsai}, C.~L., {Chen}, H.~J., {et~al.} 2009, Journal of
  Geophysical Research (Space Physics), 114, A08102,
  \dodoi{10.1029/2008JA014008}

\bibitem[{{Lin}(1997)}]{Lin97}
{Lin}, Y. 1997, \jgr, 102, 24265, \dodoi{10.1029/97JA01989}

\bibitem[{{Lin}(2002)}]{Lin02:HFA}
---. 2002, \planss, 50, 577, \dodoi{10.1016/S0032-0633(02)00037-5}

\bibitem[{{Lin} {et~al.}(1996{\natexlab{a}}){Lin}, {Lee}, \&
  {Yan}}]{Lin96:discontinuity}
{Lin}, Y., {Lee}, L.~C., \& {Yan}, M. 1996{\natexlab{a}}, \jgr, 101, 479,
  \dodoi{10.1029/95JA02985}

\bibitem[{{Lin} {et~al.}(1996{\natexlab{b}}){Lin}, {Swift}, \&
  {Lee}}]{Lin96:pressure_pulse}
{Lin}, Y., {Swift}, D.~W., \& {Lee}, L.~C. 1996{\natexlab{b}}, \jgr, 101,
  27251, \dodoi{10.1029/96JA02733}

\bibitem[{{Lipatov}(2002)}]{Lipatov}
{Lipatov}, A.~S. 2002, {The Hybrid Multiscale Simulation Technology}
  (Springer-Verlag Berlin Heidelberg)

\bibitem[{{Liu} {et~al.}(2019){Liu}, {Angelopoulos}, \& {Lu}}]{Liu19:foreshock}
{Liu}, T.~Z., {Angelopoulos}, V., \& {Lu}, S. 2019, Science Advances, 5,
  eaaw1368, \dodoi{10.1126/sciadv.aaw1368}

\bibitem[{{Liu} {et~al.}(2015){Liu}, {Turner}, {Angelopoulos}, \&
  {Omidi}}]{Liu15:foreshock}
{Liu}, Z., {Turner}, D.~L., {Angelopoulos}, V., \& {Omidi}, N. 2015, \grl, 42,
  7860, \dodoi{10.1002/2015GL065842}

\bibitem[{Matthews(1994)}]{Matthews94}
Matthews, A.~P. 1994, J. Comput. Phys., 112, 102

\bibitem[{{Maynard} {et~al.}(2007){Maynard}, {Burke}, {Ober}, {Farrugia},
  {Kucharek}, {Lester}, {Mozer}, {Russell}, \& {Siebert}}]{Maynard07}
{Maynard}, N.~C., {Burke}, W.~J., {Ober}, D.~M., {et~al.} 2007, Journal of
  Geophysical Research (Space Physics), 112, A12219,
  \dodoi{10.1029/2007JA012293}

\bibitem[{{Maynard} {et~al.}(2008){Maynard}, {Farrugia}, {Ober}, {Burke},
  {Dunlop}, {Mozer}, {R{\`e}Me}, {D{\'e}Cr{\'e}Au}, \& {Siebert}}]{Maynard08}
{Maynard}, N.~C., {Farrugia}, C.~J., {Ober}, D.~M., {et~al.} 2008, Journal of
  Geophysical Research (Space Physics), 113, A10212,
  \dodoi{10.1029/2008JA013121}

\bibitem[{{McFadden} {et~al.}(2008){McFadden}, {Carlson}, {Larson}, {Ludlam},
  {Abiad}, {Elliott}, {Turin}, {Marckwordt}, \&
  {Angelopoulos}}]{McFadden08:THEMIS}
{McFadden}, J.~P., {Carlson}, C.~W., {Larson}, D., {et~al.} 2008, \ssr, 141,
  277, \dodoi{10.1007/s11214-008-9440-2}

\bibitem[{{Medvedev} {et~al.}(1997{\natexlab{a}}){Medvedev}, {Diamond},
  {Shevchenko}, \& {Galinsky}}]{Medvedev97:prl}
{Medvedev}, M.~V., {Diamond}, P.~H., {Shevchenko}, V.~I., \& {Galinsky}, V.~L.
  1997{\natexlab{a}}, Physical Review Letters, 78, 4934,
  \dodoi{10.1103/PhysRevLett.78.4934}

\bibitem[{{Medvedev} {et~al.}(1997{\natexlab{b}}){Medvedev}, {Shevchenko},
  {Diamond}, \& {Galinsky}}]{Medvedev97:pop}
{Medvedev}, M.~V., {Shevchenko}, V.~I., {Diamond}, P.~H., \& {Galinsky}, V.~L.
  1997{\natexlab{b}}, Physics of Plasmas, 4, 1257, \dodoi{10.1063/1.872356}

\bibitem[{{Nakanotani} {et~al.}(2020){Nakanotani}, {Zank}, \&
  {Zhao}}]{Nakanotani20}
{Nakanotani}, M., {Zank}, G.~P., \& {Zhao}, L. 2020, in Journal of Physics
  Conference Series, Vol. 1620, Journal of Physics Conference Series, 012014,
  \dodoi{10.1088/1742-6596/1620/1/012014}

\bibitem[{{Neugebauer}(2006)}]{Neugebauer06}
{Neugebauer}, M. 2006, \jgr, 111, A04103, \dodoi{10.1029/2005JA011497}

\bibitem[{{Neugebauer} {et~al.}(1984){Neugebauer}, {Clay}, {Goldstein},
  {Tsurutani}, \& {Zwickl}}]{Neugebauer84}
{Neugebauer}, M., {Clay}, D.~R., {Goldstein}, B.~E., {Tsurutani}, B.~T., \&
  {Zwickl}, R.~D. 1984, \jgr, 89, 5395, \dodoi{10.1029/JA089iA07p05395}

\bibitem[{{Neukirch} {et~al.}(2020{\natexlab{a}}){Neukirch}, {Vasko},
  {Artemyev}, \& {Allanson}}]{Neukirch20}
{Neukirch}, T., {Vasko}, I.~Y., {Artemyev}, A.~V., \& {Allanson}, O.
  2020{\natexlab{a}}, \apj, 891, 86, \dodoi{10.3847/1538-4357/ab7234}

\bibitem[{{Neukirch} {et~al.}(2020{\natexlab{b}}){Neukirch}, {Wilson}, \&
  {Allanson}}]{Neukirch20:jpp}
{Neukirch}, T., {Wilson}, F., \& {Allanson}, O. 2020{\natexlab{b}}, Journal of
  Plasma Physics, 86, 825860302, \dodoi{10.1017/S0022377820000604}

\bibitem[{{Newman} {et~al.}(2020){Newman}, {Vainchtein}, \&
  {Artemyev}}]{Newman20}
{Newman}, R., {Vainchtein}, D., \& {Artemyev}, A. 2020, Solar Physics, 295,
  129, \dodoi{10.1007/s11207-020-01695-z}

\bibitem[{{Omidi}(1992)}]{Omidi92}
{Omidi}, N. 1992, \grl, 19, 1335, \dodoi{10.1029/92GL01127}

\bibitem[{{Omidi} {et~al.}(2010){Omidi}, {Eastwood}, \& {Sibeck}}]{Omidi10}
{Omidi}, N., {Eastwood}, J.~P., \& {Sibeck}, D.~G. 2010, Journal of Geophysical
  Research (Space Physics), 115, A06204, \dodoi{10.1029/2009JA014828}

\bibitem[{{Osman} {et~al.}(2011){Osman}, {Matthaeus}, {Greco}, \&
  {Servidio}}]{Osman11:solarwind}
{Osman}, K.~T., {Matthaeus}, W.~H., {Greco}, A., \& {Servidio}, S. 2011, \apjl,
  727, L11, \dodoi{10.1088/2041-8205/727/1/L11}

\bibitem[{{Osman} {et~al.}(2012){Osman}, {Matthaeus}, {Wan}, \&
  {Rappazzo}}]{Osman12:solarwind}
{Osman}, K.~T., {Matthaeus}, W.~H., {Wan}, M., \& {Rappazzo}, A.~F. 2012,
  Physical Review Letters, 108, 261102, \dodoi{10.1103/PhysRevLett.108.261102}

\bibitem[{{Phan} {et~al.}(2006){Phan}, {Gosling}, {Davis}, {Skoug},
  {{\O}ieroset}, {Lin}, {Lepping}, {McComas}, {Smith}, {Reme}, \&
  {Balogh}}]{Phan06:reconnection}
{Phan}, T.~D., {Gosling}, J.~T., {Davis}, M.~S., {et~al.} 2006, \nat, 439, 175,
  \dodoi{10.1038/nature04393}

\bibitem[{{Phan} {et~al.}(2010){Phan}, {Gosling}, {Paschmann}, {Pasma},
  {Drake}, {{\O}ieroset}, {Larson}, {Lin}, \& {Davis}}]{Phan10}
{Phan}, T.~D., {Gosling}, J.~T., {Paschmann}, G., {et~al.} 2010, \apjl, 719,
  L199, \dodoi{10.1088/2041-8205/719/2/L199}

\bibitem[{{Phan} {et~al.}(2018){Phan}, {Eastwood}, {Shay}, {Drake}, {Sonnerup},
  {Fujimoto}, {Cassak}, {{\O}ieroset}, {Burch}, {Torbert}, {Rager}, {Dorelli},
  {Gershman}, {Pollock}, {Pyakurel}, {Haggerty}, {Khotyaintsev}, {Lavraud},
  {Saito}, {Oka}, {Ergun}, {Retino}, {Le Contel}, {Argall}, {Giles}, {Moore},
  {Wilder}, {Strangeway}, {Russell}, {Lindqvist}, \& {Magnes}}]{Phan18:nature}
{Phan}, T.~D., {Eastwood}, J.~P., {Shay}, M.~A., {et~al.} 2018, \nat, 557, 202,
  \dodoi{10.1038/s41586-018-0091-5}

\bibitem[{{Phan} {et~al.}(2020){Phan}, {Bale}, {Eastwood}, {Lavraud}, {Drake},
  {Oieroset}, {Shay}, {Pulupa}, {Stevens}, {MacDowall}, {Case}, {Larson},
  {Kasper}, {Whittlesey}, {Szabo}, {Korreck}, {Bonnell}, {de Wit}, {Goetz},
  {Harvey}, {Horbury}, {Livi}, {Malaspina}, {Paulson}, {Raouafi}, \&
  {Velli}}]{Phan20}
{Phan}, T.~D., {Bale}, S.~D., {Eastwood}, J.~P., {et~al.} 2020, \apjs, 246, 34,
  \dodoi{10.3847/1538-4365/ab55ee}

\bibitem[{{Plaschke} {et~al.}(2018){Plaschke}, {Hietala}, {Archer},
  {Blanco-Cano}, {Kajdi{\v c}}, {Karlsson}, {Lee}, {Omidi}, {Palmroth},
  {Roytershteyn}, {Schmid}, {Sergeev}, \& {Sibeck}}]{Plaschke18}
{Plaschke}, F., {Hietala}, H., {Archer}, M., {et~al.} 2018, \ssr, 214, 81,
  \dodoi{10.1007/s11214-018-0516-3}

\bibitem[{{Podesta}(2017)}]{Podesta17}
{Podesta}, J.~J. 2017, \jgr, 122, 2795, \dodoi{10.1002/2016JA023629}

\bibitem[{{Podesta} \& {Roytershteyn}(2017)}]{Podesta&Roytershteyn17}
{Podesta}, J.~J., \& {Roytershteyn}, V. 2017, \jgr, 122, 6991,
  \dodoi{10.1002/2017JA024074}

\bibitem[{{Ponomaryov} {et~al.}(2019){Ponomaryov}, {Levenfish}, \&
  {Petrov}}]{Ponomaryov2019}
{Ponomaryov}, G.~A., {Levenfish}, K.~P., \& {Petrov}, A.~E. 2019, in Journal of
  Physics Conference Series, Vol. 1400, Journal of Physics Conference Series,
  022027, \dodoi{10.1088/1742-6596/1400/2/022027}

\bibitem[{{Pope} {et~al.}(2019){Pope}, {Gedalin}, \& {Balikhin}}]{Pope19}
{Pope}, S.~A., {Gedalin}, M., \& {Balikhin}, M.~A. 2019, Journal of Geophysical
  Research (Space Physics), 124, 1711, \dodoi{10.1029/2018JA026223}

\bibitem[{{Posselt} {et~al.}(2017){Posselt}, {Pavlov}, {Slane}, {Romani},
  {Bucciantini}, {Bykov}, {Kargaltsev}, {Weisskopf}, \& {Ng}}]{Posselt17}
{Posselt}, B., {Pavlov}, G.~G., {Slane}, P.~O., {et~al.} 2017, \apj, 835, 66,
  \dodoi{10.3847/1538-4357/835/1/66}

\bibitem[{{Pushkar}(2010)}]{Pushkar10}
{Pushkar}, E.~A. 2010, Fluid Dynamics, 44, 917,
  \dodoi{10.1134/S0015462809060155}

\bibitem[{{Retin{\`o}} {et~al.}(2007){Retin{\`o}}, {Sundkvist}, {Vaivads},
  {Mozer}, {Andr{\'e}}, \& {Owen}}]{Retino07:nat}
{Retin{\`o}}, A., {Sundkvist}, D., {Vaivads}, A., {et~al.} 2007, Nature
  Physics, 3, 236, \dodoi{10.1038/nphys574}

\bibitem[{{Roth} \& {Bale}(2006)}]{Roth&Bale06}
{Roth}, I., \& {Bale}, S.~D. 2006, \jgr, 111, 7, \dodoi{10.1029/2005JA011434}

\bibitem[{{Roth} {et~al.}(1996){Roth}, {de Keyser}, \& {Kuznetsova}}]{Roth96}
{Roth}, M., {de Keyser}, J., \& {Kuznetsova}, M.~M. 1996, Space Science
  Reviews, 76, 251, \dodoi{10.1007/BF00197842}

\bibitem[{Sagdeev(1966)}]{Sagdeev1966}
Sagdeev, R.~Z. 1966, Rev Plasma Phys, 4, 23

\bibitem[{{Samsonov} {et~al.}(2007){Samsonov}, {Sibeck}, \&
  {Imber}}]{Samsonov07}
{Samsonov}, A.~A., {Sibeck}, D.~G., \& {Imber}, J. 2007, Journal of Geophysical
  Research (Space Physics), 112, A12220, \dodoi{10.1029/2007JA012627}

\bibitem[{{Sergeev} {et~al.}(2006){Sergeev}, {Sormakov}, {Apatenkov},
  {Baumjohann}, {Nakamura}, {Runov}, {Mukai}, \& {Nagai}}]{Sergeev06}
{Sergeev}, V.~A., {Sormakov}, D.~A., {Apatenkov}, S.~V., {et~al.} 2006, Annales
  Geophysicae, 24, 2015

\bibitem[{{Servidio} {et~al.}(2011{\natexlab{a}}){Servidio}, {Greco},
  {Matthaeus}, {Osman}, \& {Dmitruk}}]{Servidio11}
{Servidio}, S., {Greco}, A., {Matthaeus}, W.~H., {Osman}, K.~T., \& {Dmitruk},
  P. 2011{\natexlab{a}}, \jgr, 116, 9102, \dodoi{10.1029/2011JA016569}

\bibitem[{{Servidio} {et~al.}(2015){Servidio}, {Valentini}, {Perrone}, {Greco},
  {Califano}, {Matthaeus}, \& {Veltri}}]{Servidio15}
{Servidio}, S., {Valentini}, F., {Perrone}, D., {et~al.} 2015, Journal of
  Plasma Physics, 81, 325810107, \dodoi{10.1017/S0022377814000841}

\bibitem[{{Servidio} {et~al.}(2011{\natexlab{b}}){Servidio}, {Dmitruk},
  {Greco}, {Wan}, {Donato}, {Cassak}, {Shay}, {Carbone}, \&
  {Matthaeus}}]{Servidio11:npg}
{Servidio}, S., {Dmitruk}, P., {Greco}, A., {et~al.} 2011{\natexlab{b}},
  Nonlinear Processes in Geophysics, 18, 675, \dodoi{10.5194/npg-18-675-2011}

\bibitem[{{Sironi} \& {Spitkovsky}(2011)}]{Sironi&Spitkovsky11}
{Sironi}, L., \& {Spitkovsky}, A. 2011, \apj, 741, 39,
  \dodoi{10.1088/0004-637X/741/1/39}

\bibitem[{{Smith}(1973{\natexlab{a}})}]{Smith73:I}
{Smith}, E.~J. 1973{\natexlab{a}}, \jgr, 78, 2054,
  \dodoi{10.1029/JA078i013p02054}

\bibitem[{{Smith}(1973{\natexlab{b}})}]{Smith73:II}
---. 1973{\natexlab{b}}, \jgr, 78, 2088, \dodoi{10.1029/JA078i013p02088}

\bibitem[{{S{\"o}ding} {et~al.}(2001){S{\"o}ding}, {Neubauer}, {Tsurutani},
  {Ness}, \& {Lepping}}]{Soding01}
{S{\"o}ding}, A., {Neubauer}, F.~M., {Tsurutani}, B.~T., {Ness}, N.~F., \&
  {Lepping}, R.~P. 2001, Annales Geophysicae, 19, 681

\bibitem[{{Sonnerup} \& {Cahill}(1968)}]{Sonnerup68}
{Sonnerup}, B.~U.~{\"O}., \& {Cahill}, Jr., L.~J. 1968, \jgr, 73, 1757,
  \dodoi{10.1029/JA073i005p01757}

\bibitem[{{Soucek} {et~al.}(2008){Soucek}, {Lucek}, \& {Dandouras}}]{Soucek08}
{Soucek}, J., {Lucek}, E., \& {Dandouras}, I. 2008, \jgr, 113, A04203,
  \dodoi{10.1029/2007JA012649}

\bibitem[{{Swisdak} {et~al.}(2003){Swisdak}, {Rogers}, {Drake}, \&
  {Shay}}]{Swisdak03}
{Swisdak}, M., {Rogers}, B.~N., {Drake}, J.~F., \& {Shay}, M.~A. 2003, Journal
  of Geophysical Research (Space Physics), 108, 1218,
  \dodoi{10.1029/2002JA009726}

\bibitem[{{Tessein} {et~al.}(2013){Tessein}, {Matthaeus}, {Wan}, {Osman},
  {Ruffolo}, \& {Giacalone}}]{Tessein13}
{Tessein}, J.~A., {Matthaeus}, W.~H., {Wan}, M., {et~al.} 2013, \apjl, 776, L8,
  \dodoi{10.1088/2041-8205/776/1/L8}

\bibitem[{{Treumann}(2009)}]{Treumann09}
{Treumann}, R.~A. 2009, The Astronomy and Astrophysics Review, 17, 409,
  \dodoi{10.1007/s00159-009-0024-2}

\bibitem[{{Treumann} \& {Baumjohann}(1997)}]{TreumannBaumjohann1997}
{Treumann}, R.~A., \& {Baumjohann}, W. 1997, {Advanced space plasma physics},
  \dodoi{10.1142/p020}

\bibitem[{{Tsurutani} \& {Ho}(1999)}]{Tsurutani&Ho99}
{Tsurutani}, B.~T., \& {Ho}, C.~M. 1999, Reviews of Geophysics, 37, 517,
  \dodoi{10.1029/1999RG900010}

\bibitem[{{Tsurutani} {et~al.}(2011){Tsurutani}, {Lakhina}, {Verkhoglyadova},
  {Gonzalez}, {Echer}, \& {Guarnieri}}]{Tsurutani11:jastp}
{Tsurutani}, B.~T., {Lakhina}, G.~S., {Verkhoglyadova}, O.~P., {et~al.} 2011,
  Journal of Atmospheric and Solar-Terrestrial Physics, 73, 5,
  \dodoi{10.1016/j.jastp.2010.04.001}

\bibitem[{{Turc} {et~al.}(2015){Turc}, {Fontaine}, {Savoini}, \&
  {Modolo}}]{Turc15}
{Turc}, L., {Fontaine}, D., {Savoini}, P., \& {Modolo}, R. 2015, Journal of
  Geophysical Research (Space Physics), 120, 6133, \dodoi{10.1002/2015JA021318}

\bibitem[{{Turner} {et~al.}(2013){Turner}, {Omidi}, {Sibeck}, \&
  {Angelopoulos}}]{Turner13:foreshock}
{Turner}, D.~L., {Omidi}, N., {Sibeck}, D.~G., \& {Angelopoulos}, V. 2013,
  Journal of Geophysical Research (Space Physics), 118, 1552,
  \dodoi{10.1002/jgra.50198}

\bibitem[{{Turner} {et~al.}(2018){Turner}, {Wilson}, {Liu}, {Cohen},
  {Schwartz}, {Osmane}, {Fennell}, {Clemmons}, {Blake}, {Westlake}, {Mauk},
  {Jaynes}, {Leonard}, {Baker}, {Strangeway}, {Russell}, {Gershman}, {Avanov},
  {Giles}, {Torbert}, {Broll}, {Gomez}, {Fuselier}, \&
  {Burch}}]{Turner18:nature}
{Turner}, D.~L., {Wilson}, L.~B., {Liu}, T.~Z., {et~al.} 2018, \nat, 561, 206,
  \dodoi{10.1038/s41586-018-0472-9}

\bibitem[{{Vasquez} {et~al.}(2007){Vasquez}, {Abramenko}, {Haggerty}, \&
  {Smith}}]{Vasquez07}
{Vasquez}, B.~J., {Abramenko}, V.~I., {Haggerty}, D.~K., \& {Smith}, C.~W.
  2007, Journal of Geophysical Research (Space Physics), 112, A11102,
  \dodoi{10.1029/2007JA012504}

\bibitem[{{Vasquez} \& {Hollweg}(1999)}]{Vasquez&Hollweg99}
{Vasquez}, B.~J., \& {Hollweg}, J.~V. 1999, \jgr, 104, 4681,
  \dodoi{10.1029/1998JA900090}

\bibitem[{{Vasquez} \& {Hollweg}(2001)}]{Vasquez&Hollweg01}
---. 2001, \jgr, 106, 5661, \dodoi{10.1029/2000JA000268}

\bibitem[{{Vinas} \& {Scudder}(1986)}]{Vinas&Scudder86}
{Vinas}, A.~F., \& {Scudder}, J.~D. 1986, \jgr, 91, 39,
  \dodoi{10.1029/JA091iA01p00039}

\bibitem[{{V{\"o}lk} \& {Auer}(1974)}]{Volk&Auer74}
{V{\"o}lk}, H.~J., \& {Auer}, R.-D. 1974, \jgr, 79, 40,
  \dodoi{10.1029/JA079i001p00040}

\bibitem[{{Wang} {et~al.}(2020){Wang}, {Chen}, {Bessho}, {Hesse}, {Wilson},
  {Denton}, {Ng}, {Giles}, {Torbert}, \& {Burch}}]{Wang20:discontinuity}
{Wang}, S., {Chen}, L.-J., {Bessho}, N., {et~al.} 2020, \apj, 898, 121,
  \dodoi{10.3847/1538-4357/ab9b8b}

\bibitem[{{Winske} \& {Omidi}(1996)}]{Winske96}
{Winske}, D., \& {Omidi}, N. 1996, JGR, 101, 17287, \dodoi{10.1029/96JA00982}

\bibitem[{{Yan} \& {Lee}(1996)}]{Yan&Lee95}
{Yan}, M., \& {Lee}, L.~C. 1996, \jgr, 101, 4835, \dodoi{10.1029/95JA02976}

\bibitem[{{Yee}(1966)}]{Yee1966}
{Yee}, K. 1966, IEEE Transactions on Antennas and Propagation, 14, 302,
  \dodoi{10.1109/TAP.1966.1138693}

\bibitem[{{Zank} {et~al.}(2014){Zank}, {le Roux}, {Webb}, {Dosch}, \&
  {Khabarova}}]{Zank14}
{Zank}, G.~P., {le Roux}, J.~A., {Webb}, G.~M., {Dosch}, A., \& {Khabarova}, O.
  2014, \apj, 797, 28, \dodoi{10.1088/0004-637X/797/1/28}

\bibitem[{{Zhou} {et~al.}(2020){Zhou}, {Gedalin}, {Russell}, {Angelopoulos}, \&
  {Drozdov}}]{Zhou20:shock}
{Zhou}, X., {Gedalin}, M., {Russell}, C.~T., {Angelopoulos}, V., \& {Drozdov},
  A.~Y. 2020, Journal of Geophysical Research (Space Physics), 125, e28174,
  \dodoi{10.1029/2020JA028174}

\bibitem[{{Zong} \& {Zhang}(2011)}]{Zong&Zhang11}
{Zong}, Q.-G., \& {Zhang}, H. 2011, Journal of Atmospheric and
  Solar-Terrestrial Physics, 73, 1, \dodoi{10.1016/j.jastp.2010.11.001}

\end{thebibliography}

\appendix
\section{Numerical convergence}
\label{appa}
The hybrid code ``Maximus'' has been used for a number of astrophysical tasks and demonstrated better than a 5\% total energy conservation, even better (about 1\%) momentum conservation, and exactly zero magnetic field divergence. The involved numerical scheme has two key features. A staggered grid \citep[the Yee lattice;][]{Yee1966} guarantees that ${\rm div} {\bf B} = 0$. A rather sophisticated field solver with a Total Variation Diminishing scheme \citep{leveque} effectively suppresses numerical oscillations in the shock front vicinity \citep[see][for further details]{Kropotina2019}. We showed in \citep{Bykov2019} that the results of ''Maximus`` are consistent with the results of the fully particle-in-cell code TRISTAN \citep{Buneman1993} for weak intracluster shocks.

The time step in ``Maximus'' is adjusted automatically to ensure that the fastest particles and waves cannot pass more than one grid length during one step. This is the necessary Courant condition of a numerical scheme convergence. The adaptive time step together with a sophisticated optimization of the RAM access and well-balanced MPI (Message Passing Interface, which is a communication protocol for computational systems with distributed memory) parallelization allows to substantially reduce the CPU time. Each of the fully 3D runs was competed in a quite reasonable time of about two weeks on several dozens of Intel Xeon $2.3$ GHz CPUs.

We performed a sequence of numerical test runs to adjust the grid length. For this purpose, we studied a long-time evolution of isolated RDs and found that initially thin RDs quickly approach some relatively stable thickness. This thickness and overall RD structure does not depend on the grid length as long as it is smaller than $1 d_{p, u}$. The coarser grid introduces an artificial viscosity that suppresses discontinuities' dispersive broadening (see also \cite{Ponomaryov2019} for an example of coarse grid effects in MHD). Therefore, we used grid lengths of $0.5 d_{p,u}$ or finer in the simulations.

\section {Accelerated particles trapping near an RD}
\label{appb}

An RD interacts with accelerated particles upstream of quasi-parallel shocks. The top three panels in Figure~\ref{trapping} show probability distribution functions of protons for run B. The shock front is at $x \approx 200 d_{p,u}$, and reflected particles streaming along the upstream magnetic field with positive $V_x$ are clearly seen in the foreshock. As long as they reach the RD, they are trapped and start flowing back to the shock. At the same time, their transverse ($V_y$ and $V_z$) distributions are broadened. This effect naturally leads to the pressure anisotropy of accelerated particles with $\beta_\perp > \beta_\parallel$, where $\beta_\perp$ and $\beta_\parallel$ are the ions' beta across and along the magnetic field, respectively. Such an anisotropy can cause at least two instabilities: the mirror and the Alfv\'en ion cyclotron.

For a thermal plasma with isotropic electrons and $\beta_e = 1$, the mirror instability criterion is  \citep{TreumannBaumjohann1997, Bykov2017}
\begin{equation}
\Lambda \equiv \frac{\beta_\perp}{\beta_\parallel} - 1 - \frac{1}{\beta_\perp} > 0
\label{eq:lambda}
\end{equation}

The $\Lambda(x)$ for run B is plotted in the bottom panel in Figure~\ref{trapping}. In a narrow region near the RD, $\Lambda$ is positive and compressional transverse long-wave mirror modes are expected to grow. They may become visible if the anisotropy lifetime exceeds the growth time of the mirror mode.

We also note that reflected particles in a foot of a quasi-perpendicular shock may cause mirror modes as well. However, more thorough study of the mirror instability is outside the scope of this paper.

Besides the mirror instability, an anisotropic beam of hot energetic particles in a relatively cold background plasma can lead to the Alfv\'{e}n ion cyclotron instability \citep{Cornwall1965, Gary2017}. It produces Alfv\'{e}n waves with wavelengths $\lambda > l_i$, frequencies $\omega < \Omega$, and wave vectors directed along $\bf B$. In our simulations, these modes may be associated with the wave trains that appear around an RD in the foreshock and on the way to the shock front grow up so that they nearly mask the discontinuity. As can be seen from the second row of the Figure~\ref{3timespar}, these waves satisfy $V_\perp = - V_{a,\perp}$ quite well and can be considered nearly Alfv\'{e}nic. Simulations of the interaction of an isolated RD with streaming accelerated particles confirmed that growing waves start as almost compressionless and become compressional only in the nonlinear phase. Thus, they are qualitatively consistent with the Alfv\'{e}n ion cyclotron instability. In the magnetosheath, these waves disappear, possibly due to the instability damping in a hot background plasma. More thorough analyses of these structures will be provided elsewhere.

\renewcommand\thefigure{\thesection\arabic{figure}}
\setcounter{figure}{0}
\begin{figure}[h]
    \centering
    \includegraphics{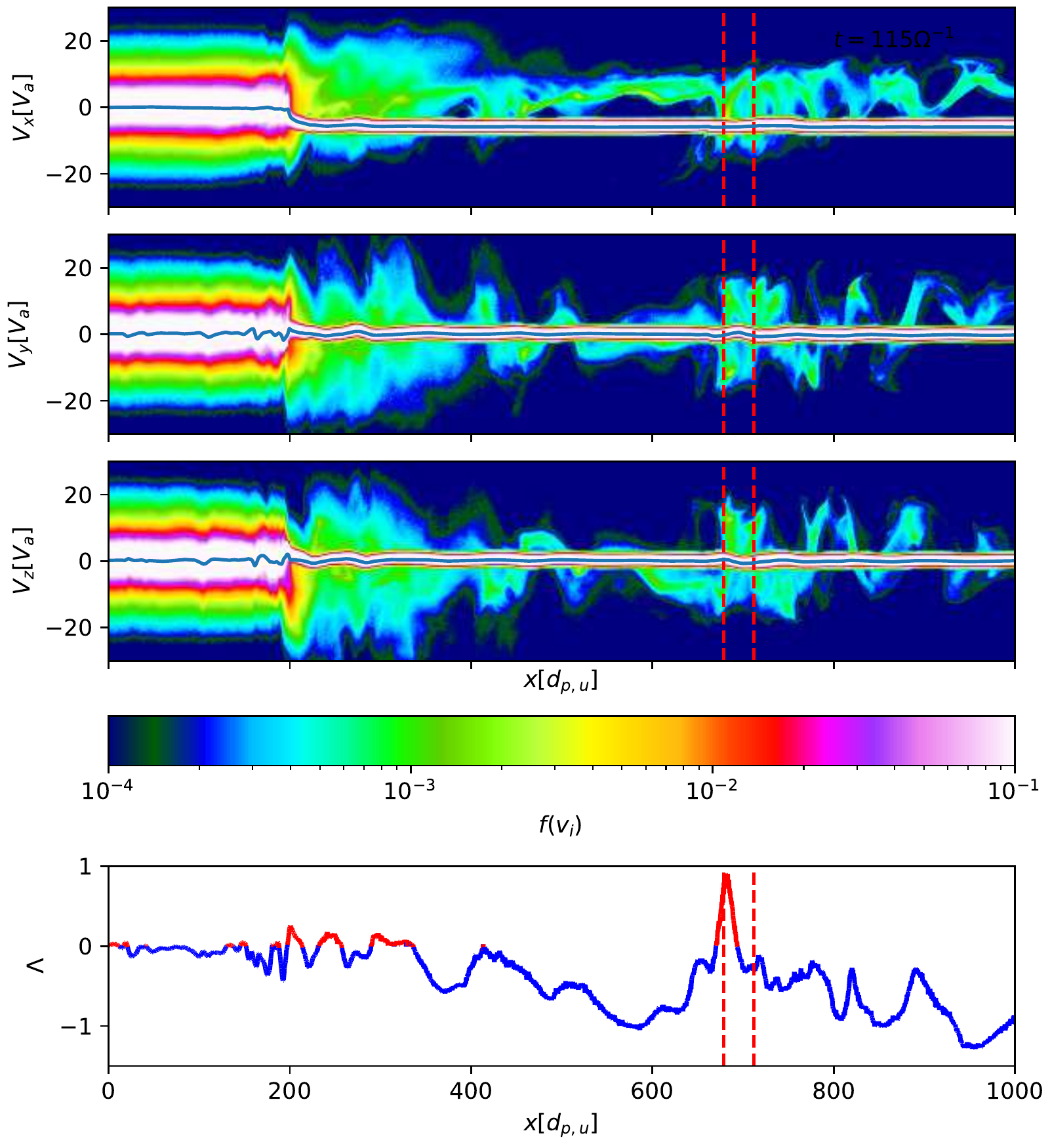}
    \caption{Protons' phase spaces and the mirror instability criterion for run B. Mean values of $V_x, \; V_y$, and $V_z$ are shown by the blue lines in the corresponding phase spaces. The RD is boxed by the red dashed lines. The sign of the mirror instability criterion $\Lambda$ is color-coded in the bottom panel.}
    \label{trapping}
\end{figure}

\end{document}